\documentclass[%
 reprint,
 amsmath,amssymb,
 aps,
 prb,
 superscriptaddress
]{revtex4-2}

\usepackage{graphicx,xcolor}
\usepackage{dcolumn}
\usepackage{bm}
\usepackage[percent]{overpic}
\usepackage{physics}
\usepackage{mathtools}
\usepackage{hyperref}
\usepackage{booktabs}
\usepackage[normalem]{ulem}

\usepackage{wasysym}
\usepackage{verbatim}

\begin{document}
\title{Classification and emergence of quantum spin liquids in chiral Rydberg models}
\author{Poetri Sonya Tarabunga}
\affiliation{The Abdus Salam International Centre for Theoretical Physics (ICTP), Strada Costiera 11, 34151 Trieste,
Italy}
\affiliation{International School for Advanced Studies (SISSA), via Bonomea 265, 34136 Trieste, Italy}
\affiliation{INFN, Sezione di Trieste, Via Valerio 2, 34127 Trieste, Italy}

\author{Giuliano Giudici}
\affiliation{Department of Physics and Arnold Sommerfeld Center for Theoretical Physics (ASC), Ludwig-Maximilians-Universit\"at M\"unchen, Theresienstra\ss e 37, D-80333 M\"unchen, Germany}
\affiliation{Munich Center for Quantum Science and Technology (MCQST), Schellingstra\ss e 4, D-80799 M\"unchen, Germany}

\author{Titas Chanda}
\affiliation{The Abdus Salam International Centre for Theoretical Physics (ICTP), Strada Costiera 11, 34151 Trieste,
Italy}

\author{Marcello Dalmonte}
\affiliation{The Abdus Salam International Centre for Theoretical Physics (ICTP), Strada Costiera 11, 34151 Trieste,
Italy}
\affiliation{International School for Advanced Studies (SISSA), via Bonomea 265, 34136 Trieste, Italy}

\date{\today}

\begin{abstract}
We investigate the nature of quantum phases arising in chiral interacting Hamiltonians recently realized in Rydberg atom arrays. We classify all possible fermionic chiral spin liquids with $\mathrm{U}(1)$ global symmetry using parton construction on the honeycomb lattice. The resulting classification includes six distinct classes of gapped quantum spin liquids: the corresponding variational wavefunctions obtained from two of these classes accurately describe the Rydberg many-body ground state at $1/2$ and $1/4$ particle density. Complementing this analysis with tensor network simulations, we conclude that both particle filling sectors host a spin liquid with the same topological order of a $\nu$\,$=$\,$1/2$ fractional quantum Hall effect.
At density $1/2$, our results clarify the phase diagram of the model, while at density $1/4$, they provide an explicit construction of the ground state wavefunction with almost unit overlap with the microscopic one. 
These findings pave the way to the use of parton wavefunctions to guide the discovery of quantum spin liquids in chiral Rydberg models.
\end{abstract}
\maketitle

\section{Introduction} There is presently considerable interest in studying strongly correlated phases of matter in synthetic quantum systems based on Rydberg atom arrays~\cite{gross2017quantum,browaeys_2020}. Stimulated by early experiments realizing symmetry-protected topological phases in one dimension~\cite{SSHRyd2019}, these platforms are now able to realize frustrated Hamiltonian dynamics in two dimensions~\cite{scholl2021quantum,ebadi2021quantum,Semeghini2021}, thus providing unparalleled opportunities to realize quantum spin liquids (QSLs) -- elusive, exotic states of matter which have captivated the attention of physicists for decades \cite{Savary_2016,Wen_2017,lacroix2011,moessner_moore_2021}. One route to access QSLs is based on the realization of frustrated Ising models in the so-called frozen gas regime~\cite{lukin2001dipole,browaeys_2020,schauss2018quantum}: several theoretical works have proposed different realistic scenarios for both gapped and gapless phases of matter \cite{Samajdar2021,Verresen2021,giudici2022,Giudice_2022}, with pioneering experiments already reporting evidence for deconfinement \cite{Semeghini2021}. These models resemble closely situations investigated in the context of quantum dimer models \cite{moessner2001ising}, providing direct link between gauge theories and experimental settings \cite{glaetzle2014quantum,Tarabunga_2022,Samajdar2022,Cheng2023}. 

Over the last two years, a new route has been paved in a very different experimental regime, where the dynamics solely takes place within the Rydberg subspace. The resulting Hamiltonians naturally feature various forms of chiral multi-body interactions \cite{peter_2015,Weber_2018,Weber_2022,Ohler_2022}, which have already been experimentally demonstrated~\cite{Lienhard_2020}. These classes of dynamics differ fundamentally from traditional Ising- and Heisenberg-type frustrated magnets and, while very promising since they display chiral terms, it is presently not even clear what classes of quantum spin liquids these can stabilize, and in which parameter regimes they might be observed. 

In this work, we provide a general framework to describe chiral spin liquids (CSLs) in Rydberg atom honeycomb arrays. This framework is based on a systematic CSLs classification ~\cite{Wen_2002,Wen_2002_2,Bieri_2016} using a fermionic spinon construction ~\cite{Baskaran_1988,Baskaran_1989} that yields Gutzwiller-projected parton wavefunction {\it ansätze} for the many-body ground state. The resulting classification differs substantially from those of Heisenberg-type regimes: it rules out the possibility of gapless Dirac spectrum,  while predicting several, distinct topological phases. 

Combining variational wavefunctions obtained from the classification with exact diagonalization (ED) methods, we demonstrate that the former capture the intermediate liquid regime of the chiral Rydberg model at both 1/2~\cite{Ohler_2022} and 1/4 density~\cite{Weber_2022}, which - surprisingly - encode the same form of topological order: a two-fold ground state degeneracy and a fractionalized Chern number $C$\,$=$\,$1/2$ per state. These two CSLs represent two distinct phases characterized by different projective representations of the lattice symmetries in the underlying fermionic spinon space. Remarkably, the CSL at 1/2 density is a new phase which corresponds to integer filling of the single-particle band, thereby representing an interaction-driven  
 topological phase generated from a trivial band insulator~\cite{footnote1}, setting an open quest recently put forward in Ref.~\cite{Ohler_2022}. We then corroborate the topological character of this phase by computing the topological correction to the area-law scaling of entanglement entropy~\cite{kitaev2006,levin2006}, which is consistent with a CSL ground state, and by analyzing the pattern of currents at the edges of a cylinder using the density-matrix-renormalization-group (DMRG)~\cite{white_prl_1992,white_prb_1993,schollwock_aop_2011, Orus_aop_2014}, which shows substantial counter-propagating nearest-neighbor edge currents, offering a simple mean for experimental detection. In the 1/4-density case, our results allow to frame the recent observation of CSL~\cite{Weber_2022} within a rigorous classification, as well as providing a genuine understanding of the system wavefunction.

The rest of the paper is structured as follows. In Sec.~\ref{sec:model}, we introduce the model under study and give a brief overview of previous works on its phase diagram. In Sec.~\ref{sec:classification}, we present the classification of CSL on the honeycomb lattice with appropriate symmetries of the Rydberg model. The microscopic wavefunctions resulting from the classification are then compared with the ground-state wavefunction of the model in Sec.~\ref{sec:overlaps}. Here, we show that the overlaps are significant in the intermediate phase, strongly indicating that the phase is a CSL. In Sec.~\ref{sec:spectra}, Sec.~\ref{sec:tee} and Sec.~\ref{sec:currents}, we show additional numerical results on excitation spectra, topological entanglement entropy, and chiral edge currents, respectively, which further confirm the CSL nature of the intermediate phase. Then, in Sec. \ref{sec:new_phase}, we discuss a putative intermediate phase intervening between the CSL and the BEC phase. Finally we conclude in Sec.~\ref{sec:concl}.

\begin{figure*}[htb]
    \centering
\includegraphics[width=\linewidth]{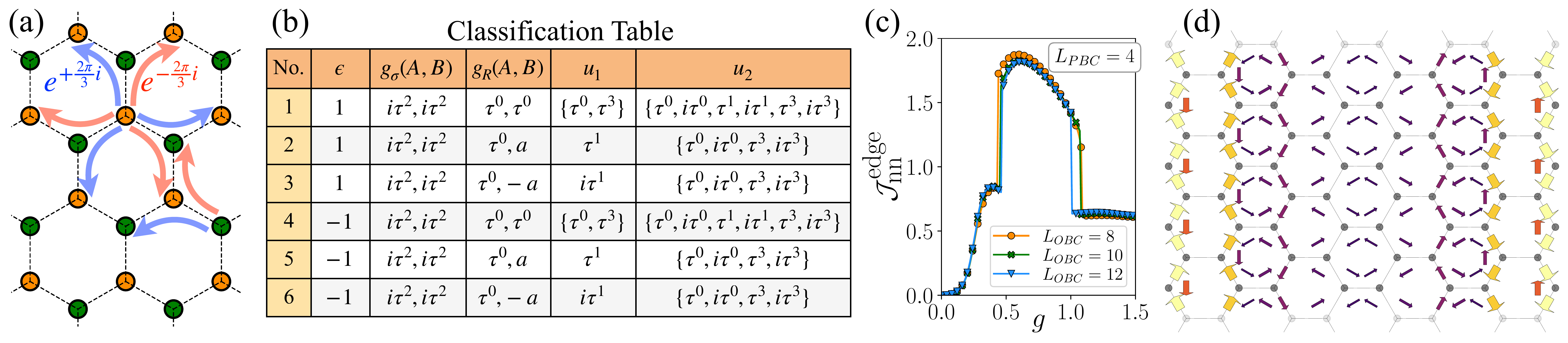}
    \caption{(a) Schematics of the model on the honeycomb lattice. The signs of the complex phase in the next-nearest-neightbor (NNN) hoppings are indicated by the colored arrows. (b) Mean-field ans\"atze of chiral spin liquids on the honeycomb lattice with broken time-reversal and reflection symmetries. $\epsilon$ indicates whether the unit-cell is doubled in the spinon space. $g_\sigma(A/B)$ and $g_R(A/B)$ are projective representations of reflection and rotation symmetries, respectively, with $a$\,$=$\,$e^{ i 2\pi /3 \tau^3}$. $u_1$ and $u_2$ are mean-field amplitudes at NN and NNN links, respectively. The CSL phase at 1/2  and 1/4-density are captured by ansatz no.~1 and 4, respectively.
    (c) NN edge currents for cylinders with periodic width $L_{\text{PBC}}$\,$=$\,$4$ (i.e., 8 lattice sites). The edge currents are substantially larger in the intermediate phase compared to the neighboring phases. (d) The current profile at $g=0.74$ for a cylinder with periodic width $L_{\text{PBC}}$\,$=$\,$4$ and length $L_{\text{OBC}}$\,$=$\,$8$. Widths of the arrows are proportional to the current values and directions denote the current directions.  Large counter-propagating NN currents are observed only at the edges, while they vanish in the bulk.}
    \label{fig:schematic}
\end{figure*}

\section{Model Hamiltonian and phase diagram} \label{sec:model}

We consider a system with atoms arranged on a honeycomb lattice. We assume optical control on the three Rydberg states $|0\rangle$, $| + \rangle$\,$=$\,$a^\dagger | 0 \rangle$, and $| - \rangle$\,$=$\,$b^\dagger | 0 \rangle$, where the state $|0\rangle$ has no excitations, and the states $|+\rangle$ and $|-\rangle$ (which belong to the same Rydberg manifold and differ, e.g., by their total angular momentum) encode two species of hard-core bosons created by the operators $a^\dagger$ and $b^\dagger$, respectively. Atomic motion is irrelevant on typical experimental timescales: below, we focus on the dynamics of the Rydberg excitations.
The model Hamiltonian is \cite{peter_2015,Weber_2018}:

\begin{align} 
    \nonumber 
    H_0 = &\sum_{i \neq j} \begin{pmatrix} a_i \\ b_i \end{pmatrix}^\dagger \begin{pmatrix}
         -t^a_{ij} && w_{ij} e^{-i 2\phi_{ij}} \\ w_{ij} e^{i 2\phi_{ij}} && -t^b_{ij} \end{pmatrix} \begin{pmatrix} a_j \\ b_j \end{pmatrix} \\
         &+ \frac{\mu}{2} \sum_i (n^a_i- n^b_i). \label{eq:fullModel}
\end{align}

The first term represents hopping processes (of excitations) between different sites, with the real hopping conserving the internal state, and the complex hopping resulting in a change of the internal state. The amplitudes for real and complex hoppings are given by $t^a_{ij}$, $t^b_{ij}$, and $w_{ij} e^{\pm i 2\phi_{ij}}$ respectively, with $\phi_{ij}$ being the polar angle between the two Rydberg atoms on the sites $i$ and $j$. All the amplitudes scale as $1/r_{ij}^3$. The second term represents the energy difference between the two internal states, with $n^a_i$ and $n^b_i$ being the particle number operators for the $| + \rangle$ and $| - \rangle$ states, respectively.

Here, we focus on the regime $\mu \gg t^a_{ij},t^b_{ij},w$, in which case the internal state $| + \rangle$ can be adiabatically eliminated. We further make an approximation by considering only nearest-neighbor (NN) interactions in Eq.\,\eqref{eq:fullModel}, with NN hopping amplitudes $t^a, t^b,$ and $w e^{\pm i 2\phi_{ij}}$. A more detailed discussion of the validity of this approximation can be found in App. \ref{app:approx}. At leading order, the effective Hamiltonian is given by \cite{Lienhard_2020}
\begin{align} 
\nonumber 
    H = &-J \sum_{\langle ij \rangle} b^\dagger_j b_i - 2 g J \sum_{\langle \langle ij \rangle \rangle} b^\dagger_j b_i e^{\pm 2\pi i/3} (1-n_{ij}) + \text{h.c.} \\
    &+ 4 g J \sum_{\langle ij \rangle} n_i n_j, \label{eq:model}
\end{align}
where $J$\,$=$\,$t^b$ and $g$\,$=$\,$w^2/(2\mu)$. The complex phases $e^{\pm 2\pi i/3}$ in the next-nearest-neighbor (NNN) hopping are illustrated in Fig. \ref{fig:schematic}a. The NNN hoppings explicitly break time-reversal and reflection symmetries, but preserves their combination. The Hamiltonian has U(1) symmetry related to particle-number conservation. Note that, in the language of spins, the U(1) symmetry corresponds to spin-rotation symmetry around the $z$-axis. Hereafter, we set the energy scale to $J$\,$=$\,$1$. 

The phase diagram of the model at 1/2-density has been studied in Ref.~\cite{Ohler_2022}, where three different phases were found for $g$\,$\geq$\,$0$. For $g$\,$\lesssim$\,$0.4$, the phase is a Bose-Einstein condensate (BEC) \cite{footnote5}, while for $g$\,$\gtrsim$\,$0.9$ the phase exhibits spiral or $120^\circ$ spin order. The intermediate phase between $0.4$\,$\lesssim$\,$g$\,$\lesssim$\,$0.9$, shows no clear order, and it is believed to be a candidate for a spin liquid. However, its true nature remains unclear, also due to hard-to-interpret spectral properties.

At 1/4-density, Ref.~\cite{Weber_2022} investigated the full model in Eq.\,\eqref{eq:fullModel} and provided clear numerical evidence for a fractional Chern insulating state: that included ground state degeneracy and Chern number compatible with a $\nu$\,$=$\,$1/2$ bosonic Fractional Quantum Hall (FQH) state. Building on such numerical understanding, we will show below how that reflects into a very clear ansatz for the system wavefunction, informed by our classification.

\section{Classification and variational wavefunctions from parton construction} \label{sec:classification}

In order to construct a spin liquid wavefunction, a method based on fermionic representation of spins have been introduced in \cite{Baskaran_1988, Baskaran_1989}. The main idea is to fractionalize the spin-$1/2$ operators into fermionic spinon operators as $S^a$\,$=$\,$\frac{1}{2} f_i^\dagger \sigma^a_{ij} f_j$, where $\sigma^a$ are Pauli matrices, with the constraint of one spinon per site. It is convenient to introduce a two-component spinor $\Psi$\,$=$\,$(f_{\uparrow}  \ f_{\downarrow}^\dagger)^T$. Directly rewriting the spins in terms of spinons gives rise to quartic spinon interactions, which after performing mean-field approximation, leads to a quadratic spinon Hamiltonian
\begin{equation} \label{eq:mf_Ham}
    H_{MF} = \sum_{ij} \Psi^\dagger_i u_{ij} \Psi_j + \text{h.c.},
\end{equation}
where $u_{ij}$ are the mean-field amplitudes. The spinon interactions include hopping and pairing terms. The mean-field Hamiltonian is invariant under global spin rotation around the $z$-axis \cite{Dodds_2013,Reuther_2014}. The matrices $u_{ij}$ can be written as $u_{ij}$\,$=$\,$u_{ij}^\mu \sigma^\mu$, where $(\sigma^\mu)$\,$=$\,$(i\tau^0,\tau^a)$, $u_{ij}^\mu$ are complex parameters and $\tau^a$ are Pauli matrices. Real $u_{ij}^\mu$ correspond to singlet terms, while imaginary $u_{ij}^\mu$ correspond to triplet terms~\cite{footnote2}. Different mean-field ans\"atze are described by different (gauge-inequivalent) $\{u_{ij}\}$ on the links of the lattice. Finally, a physical spin state is obtained by applying Gutzwiller projection $|\psi \rangle$\,$=$\,$P_G |\psi_{MF} \rangle$, with $P_G$\,$=$\,$\prod_i n_i (2-n_i)$, to the mean-field ground state $|\psi_{MF} \rangle$. 

A method to systematically classify all possible spin liquids within this parton construction has been introduced by Wen \cite{Wen_2002,Wen_2002_2,wen2004}, based on projective symmetry groups (PSG). It has subsequently been extended to classify spin liquid phases in the absence of time-reversal (i.e., CSL) \cite{Bieri_2016} and SU(2) spin-rotation \cite{Dodds_2013,Reuther_2014} symmetries. Here, we are interested in a CSL which breaks time-reversal and reflection symmetries but preserve their combination, and which preserves U(1) spin-rotation symmetry. Such chiral mean-field states are stable beyond mean-field treatment, as the mean-field gauge fluctuations are gapped out by the Chern-Simons mechanism \cite{Wen_1989}. 

The PSG classification of symmetric spin liquids on the honeycomb lattice has been worked out in \cite{Lu_2011}, where 160 different algebraic PSG's are found. In the absence of time-reversal symmetry, we do not need to specify the SU(2) representation of the time-reversal operation \cite{Bieri_2016}. Thus, we find that the number of different classes of algebraic PSG is reduced to 24 (see App. \ref{app:psg}). Each PSG is characterized by the representations of reflection, $g_\sigma(A,B)$, and $\pi /3$ rotation, $g_R(A,B)$, for each sublattice $A$ and $B$ (for more technical details, see App. \ref{app:psg}). 

Given the model that we study, we focus on those ans\"atze that have nonzero mean-field amplitudes on the NN and NNN links. This leaves 6 distinct ans\"atze, which are listed in the Fig. \ref{fig:schematic}b. The last 2 columns indicate the symmetry-allowed terms in the mean-field Hamiltonian on the NN and NNN links \cite{footnote3}. Their amplitudes are taken as variational parameters in the following section.

Note that if the ans\"atze are restricted to NN interactions, the mean-field states are gapless with Dirac spectrum (in particular, ansatz no.~1 corresponds to the SU(2) algebraic spin liquid (ASL) state discussed in \cite{Hermele_2007}, or equivalently the u-RVB state discussed in \cite{Lu_2011}). Thus, the resulting states after Gutzwiller projection describe a Dirac spin liquid (DSL). 
However, this DSL ansatz submanifold preserves time-reversal, which is explicitly broken by our Hamiltonian. 
This excludes the possibility of a DSL being stabilized in chiral systems such as our model.

\section{Overlaps with Gutzwiller-projected parton wavefunctions} \label{sec:overlaps}
To determine whether the intermediate phase of the model in Eq.~\eqref{eq:model} is described by one of the ans\"atze above, we optimize the variational parameters by maximizing the overlap of the exact ground state of the Hamiltonian with the wavefunction ansatz, for each of the 6 ans\"atze.  
The optimization of the overlap is performed using the Nelder-Mead optimization method, implemented in MATLAB. The optimization is performed on a 16-site cluster at $g$\,$=$\,$0.7$. We find that the ansatz with the largest overlap with the ground state at 1/2-density is ansatz no.~1, characterized by non-zero real triplet NN hopping parameter $\tau^0$, imaginary singlet NNN hopping parameter $i\tau^0$, and imaginary triplet NNN hopping parameter $i\tau^3$. The corresponding amplitudes are $u_2^{i\tau^0}/u_1^{\tau^0}=-0.31$ and $u_2^{i\tau^3}/u_1^{\tau^0}=-0.1$. We have also checked that the optimal parameters do not differ much from the optimal parameters on the smaller clusters.

\begin{figure} [t]
    \centering
    \includegraphics[scale=0.31]{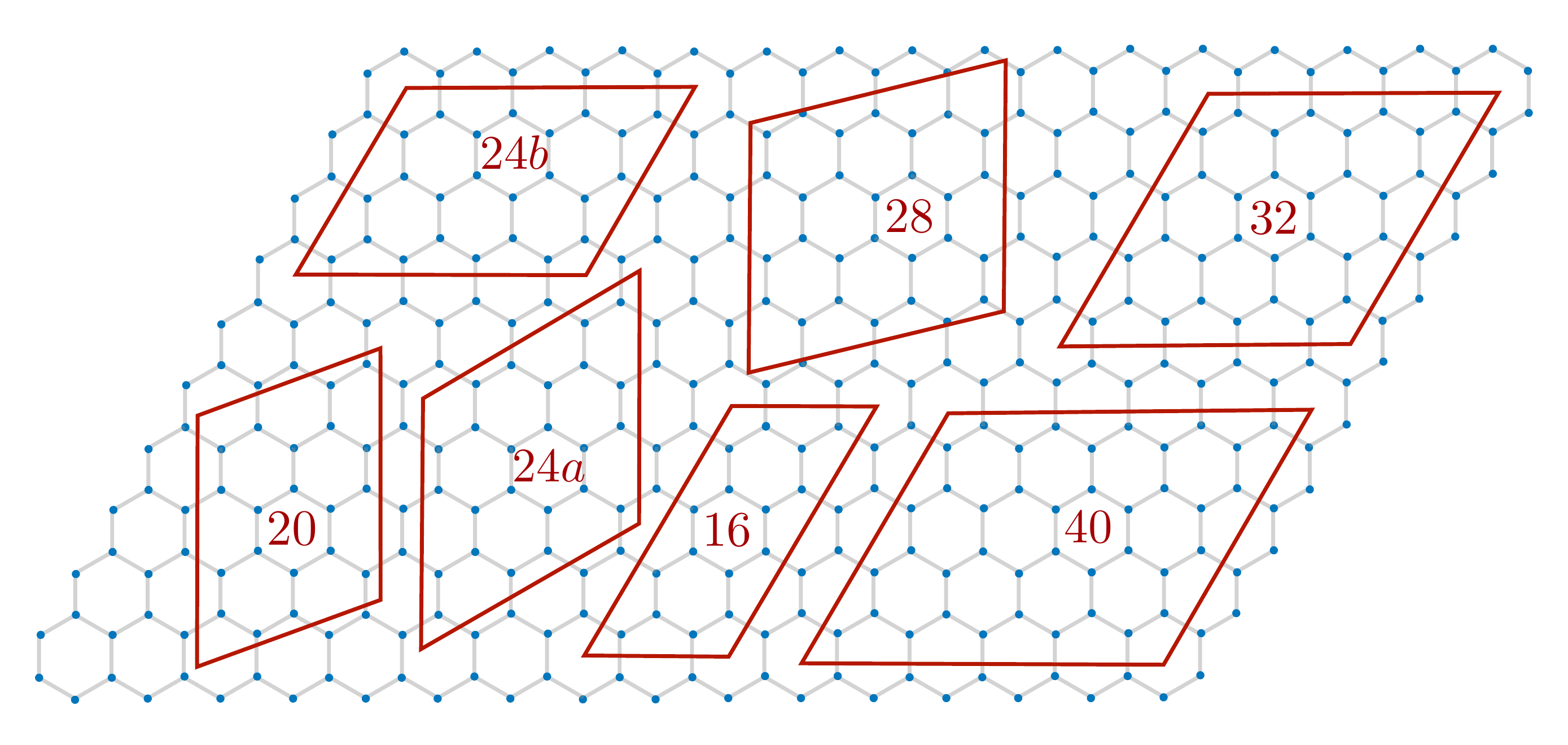}
    \caption{Periodic clusters used for the exact diagonalization calculations.}
     \label{fig:clusters}
\end{figure}

\begin{figure} 
    \centering
    \begin{overpic}[width=0.45\linewidth]{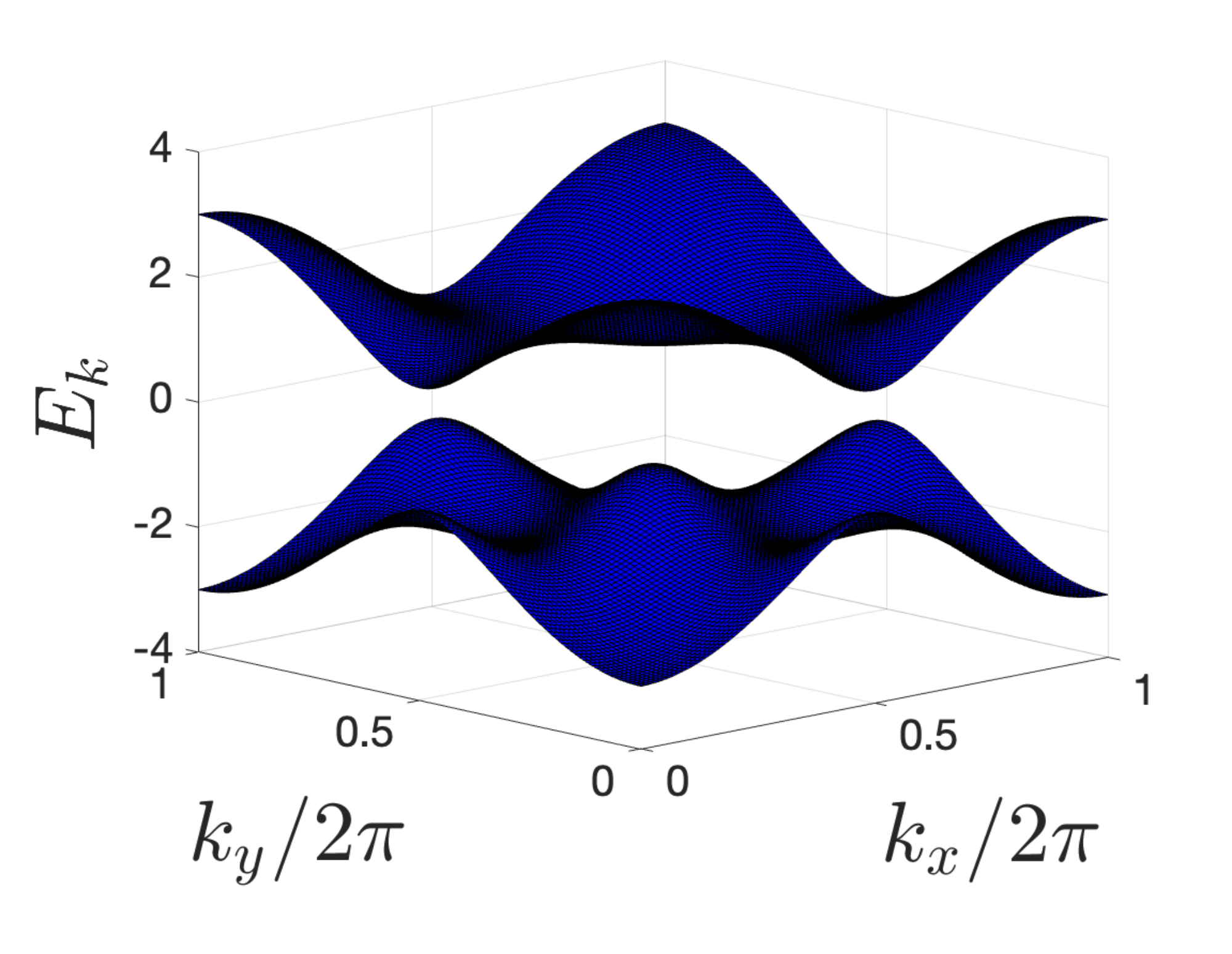}
    \put (45,80) {{\textbf{(a)}}}
    \end{overpic}
    \begin{overpic}[width=0.45\linewidth]{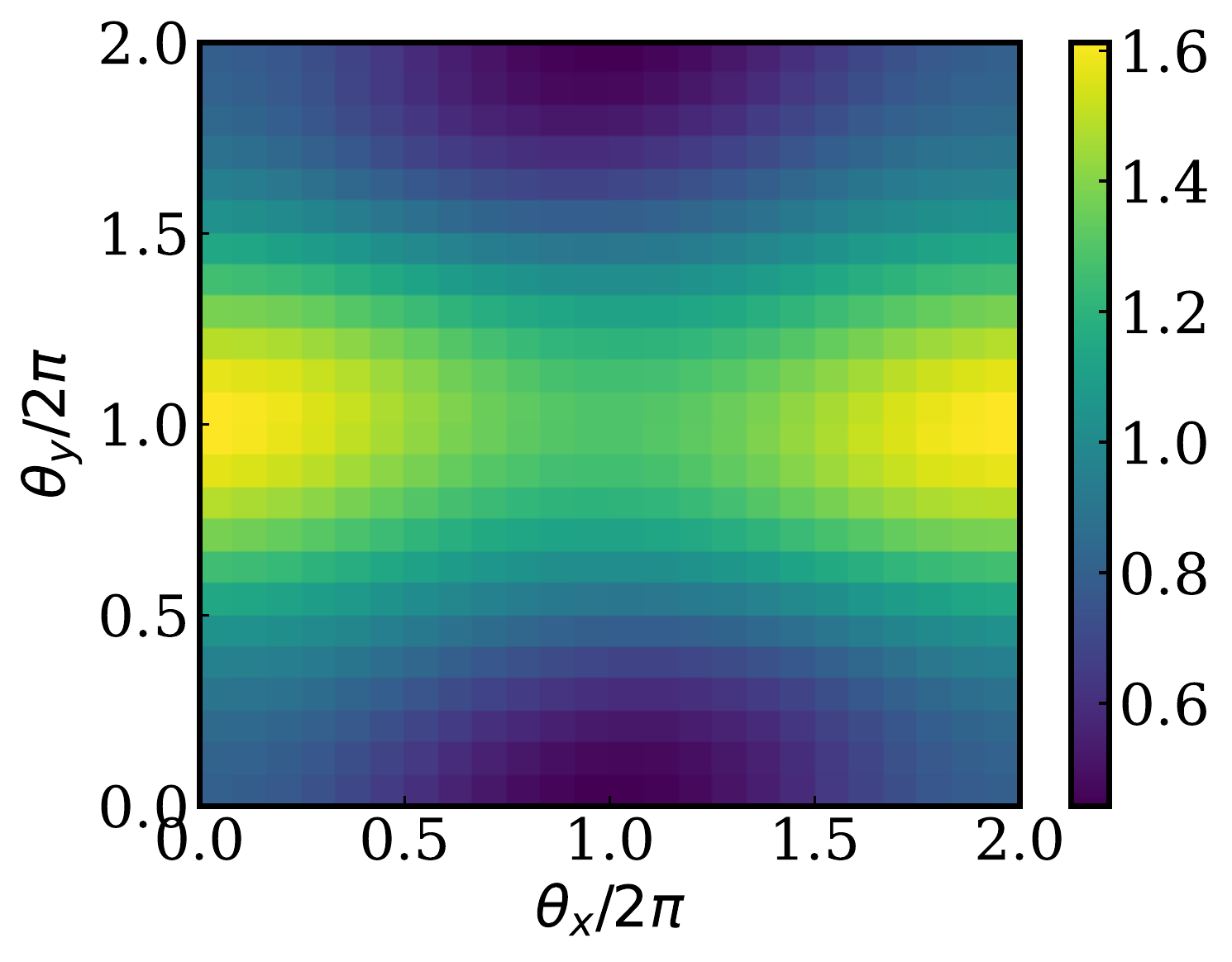}
    \put (45,80) {{\textbf{(b)}}}
    \end{overpic}
    \caption{(a) Spinon bandstructure and (b) Berry curvature on the cluster 24b (normalized with the average) of the optimal ansatz at $g=0.7$ for $n=1/2$. The parameters are  $u_2^{i\tau^0}/u_1^{\tau^0}=-0.31$ and $u_2^{i\tau^3}/u_1^{\tau^0}=-0.1$. }
    \label{fig:bandstructure}
\end{figure}

The spinon band structure with the optimal parameters are shown in Fig. \ref{fig:bandstructure}a. Note that each band is spin-degenerate.
With one fermion per site, the mean-field ground state is obtained by filling all single-particle orbitals in the lower band, for both spin-up and spin-down orbitals. There is a finite energy gap to the valence band, resulting in a fully gapped state. The Gutzwiller projection of such an ansatz has been shown to yield a topological CSL \cite{Wen_1989,Zhang_2011,Zhang_2013}, which is a lattice analogue of $\nu$\,$=$\,$1/2$ FQH Laughlin state \cite{laughlin1983}. The topological nature of such CSL is manifested by the two-fold topological degeneracy of states on a torus. These degenerate states can be constructed by threading fluxes along the non-contractible loops on a torus \cite{Wei_2015}, which can be implemented by twisting the boundary conditions of the spinons, $\Psi\rightarrow e^{i\theta/2}\Psi$. Although there are four states that can be constructed with $\theta_x,\theta_y$\,$\in$\,$\{0,\pi \}$, they only span a 2-dimensional space, resulting in two topological states. We verify this numerically by computing the overlap matrix for the four states, defined as $O_{ij}$\,$=$\,$\langle \psi_j | \psi_i \rangle$. We find that the rank of the overlap matrix is 2, within a numerical accuracy on the order of $10^{-2}$. The two independent states are then constructed from the eigenvectors of the overlap matrix with non-zero eigenvalues. 

Furthermore, we computed the many-body Chern number, a topological invariant that characterizes the topologically nontrivial phases of matter~\cite{thouless1998topological}. It can be computed by twisting the boundary condition by angles $\theta_{x,y}$ in the $x,y$ direction, and is given by the integral of the Berry curvature over the twist space: 
\begin{align} \label{eq:chern}
\nonumber
C= \frac{1}{2\pi i} \int_0^{2\pi} \int_0^{2\pi } d\theta_x d\theta_y ( &\langle \partial_{\theta_x} \Psi(\theta)^* | \partial_{\theta_y} \Psi(\theta) \rangle \\
&- \langle \partial_{\theta_y} \Psi(\theta)^* | \partial_{\theta_x} \Psi(\theta) \rangle ),  
\end{align}
where $|\Psi(\theta)\rangle=|\Psi(\theta_x,\theta_y)\rangle$ is the ground state wavefunction with twist angles $\theta_x$ and $\theta_y$. To compute Eq. \eqref{eq:chern} numerically, we discretize the twist space into $D\times D$ mesh and sum the discretized Berry curvature. We have obtained $C=2$, with the Berry curvature for $D=24$ is shown in Fig. \ref{fig:bandstructure}b. Note that the twist $0\leq\theta_{x,y}\leq2\pi$ for the fermionic spinon operators corresponds to 0 to $4\pi$ twist for the spin operators, and therefore, the result must be divided by 4. Thus, the Chern number of the spin wavefunction is fractionalized $C=1/2$ per state. 
The Chern number, along with the two-fold degeneracy, are consistent with the properties of $\nu$\,$=$\,$1/2$ bosonic Laughlin state.

Having established the FQH nature of the ansatz, we next compute the overlap of the ground state of the Hamiltonian with the two topological states  \cite{Wietek_2015,Wietek_2017},
\begin{equation}
\mathcal{O}^{ED}_{GW}=\sqrt{|\langle \psi_{ED} | \psi_{GW}^{1} \rangle|^2 + |\langle \psi_{ED} | \psi_{GW}^{2} \rangle|^2 }.    
\end{equation}
where $|\psi_{ED} \rangle$ is the ground state obtained with ED and $|\psi_{GW}^{1}\rangle$ and $|\psi_{GW}^{2}\rangle$ are the two topological states. All the ED clusters that are considered are depicted in Fig.\ref{fig:clusters}. We impose periodic boundary conditions, and the ED calculations are performed exploiting translational symmetry. The cluster 24a and the 32-site cluster has sixfold rotational symmetry, which we also exploit.

\begin{figure}
    \includegraphics[scale=0.43]{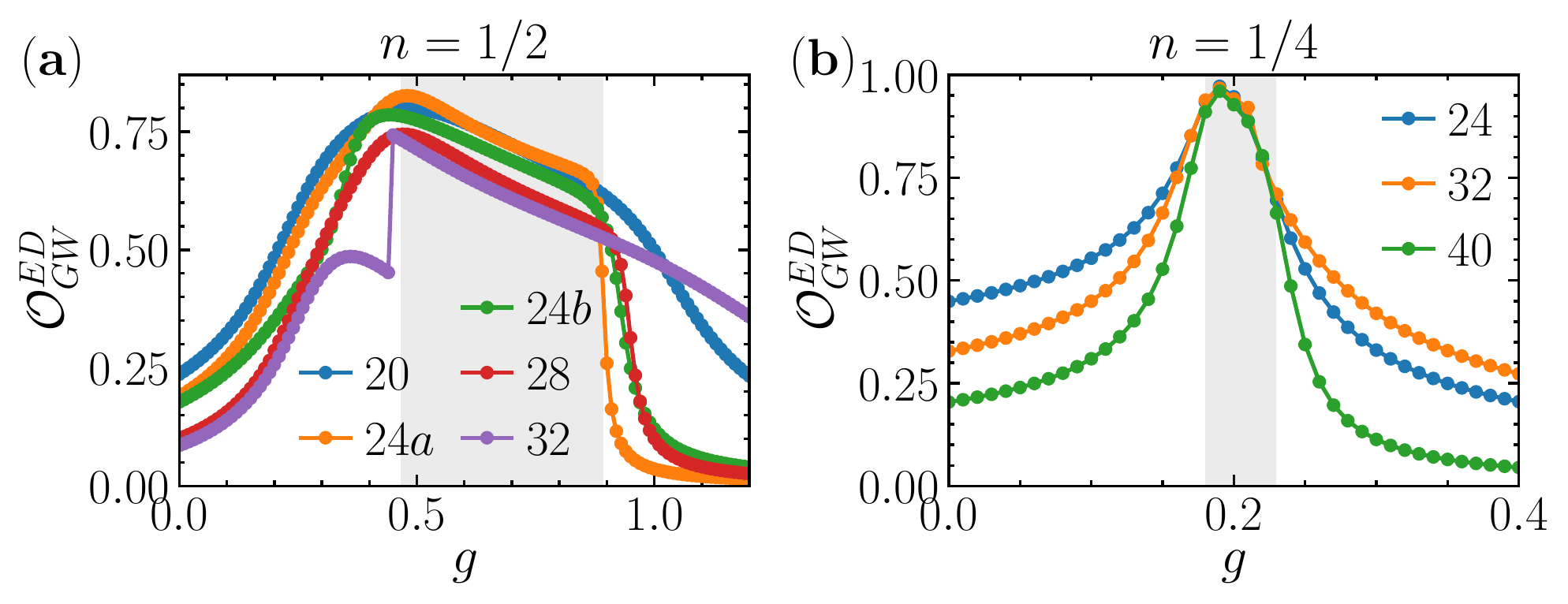}
    \vspace*{-6mm}
    \caption{Overlaps $\mathcal{O}^{ED}_{GW}$ between the exact ground states with (a) ansatz no.~1 at 1/2-density and (b) ansatz no.~4 at 1/4-density. The shaded region denotes the intermediate phase which we show to be a CSL.}
    \label{fig:overlaps}
\end{figure}

The results are shown in Fig. \ref{fig:overlaps}a for different system sizes. It can be seen that the overlap remains large in the middle phase with increasing system size, indicating that the ground state is strongly related to the topological states. Interestingly, on a 32-site cluster, we find that the ground state in the intermediate phase is not in the rotation-neutral sector. Specifically, if we take the operator $R_{\pi/3}$ which generates a $\pi/3$ rotation around the center of a honeycomb plaquette, then $R_{\pi/3} | \psi_{ED} \rangle$\,$=$\,$ e^{-2\pi i/3} | \psi_{ED} \rangle$. This can be seen in Fig. \ref{fig:overlaps}a as a discontinuous jump in the overlap at $g \approx 0.45$, as the transition from the BEC phase becomes a level crossing between different rotation sectors. Remarkably, one of the two topological states is in the same nontrivial rotation sector as the ground state of the Hamiltonian, i.e., the eigenvalue of $R_{\pi/3}$ is $e^{-2\pi i/3}$. This nontrivial observation strongly supports the hypothesis that the intermediate phase is described by the wavefunction ansatz.

At this point, it is worth emphasizing that the CSL phase found at 1/2-density represents a novel phase that has not been previously identified.
Notably, this CSL has a dispersive band (see Fig. \ref{fig:bandstructure}a), distinguishing it from the previously observed CSLs at 1/4-density \cite{Weber_2022,wang2011} and 1/8-density \cite{wang2011}, that possesses topological flat bands. As such, this CSL is inevitably missed by previous approaches relying on the identification of flat bands \cite{Weber_2022,wang2011}. 
Moreover, its dispersive band significantly affects the physical properties on finite-size clusters, thus hindering the identification of the true CSL nature in previous study \cite{Ohler_2022}. 
Our hybrid approach, combining theoretical PSG classification and numerical optimization, thus showcases its effectiveness by successfully identifying the CSL phase even in the presence of strong finite-size effects.

To compare with the 1/2-density case, we performed the same parameter optimization procedure for the 1/4-density case. In \cite{Weber_2022}, it was shown that a CSL emerges at 1/4-density for the full model in Eq.\,\eqref{eq:fullModel}. For the effective model in Eq.\,\eqref{eq:model}, we found that the CSL phase emerges in a narrow range around $g$\,$=$\,$0.2$. For 1/4-density, a gapped phase can be obtained within the parton construction when the mean-field ansatz has a doubled unit-cell. We obtain large overlaps with the ansatz no. 4 at small-size clusters in the CSL phase. Fig. \ref{fig:overlaps}b shows the overlaps for 1/4-density with the optimized parameter for different system sizes. We found that the overlap remains huge in the CSL phase, reaching 0.96 at the largest system size we considered, $L$\,$=$\,$40$. 

\begin{figure}
    \centering
    \centering
    \includegraphics[scale=0.43]{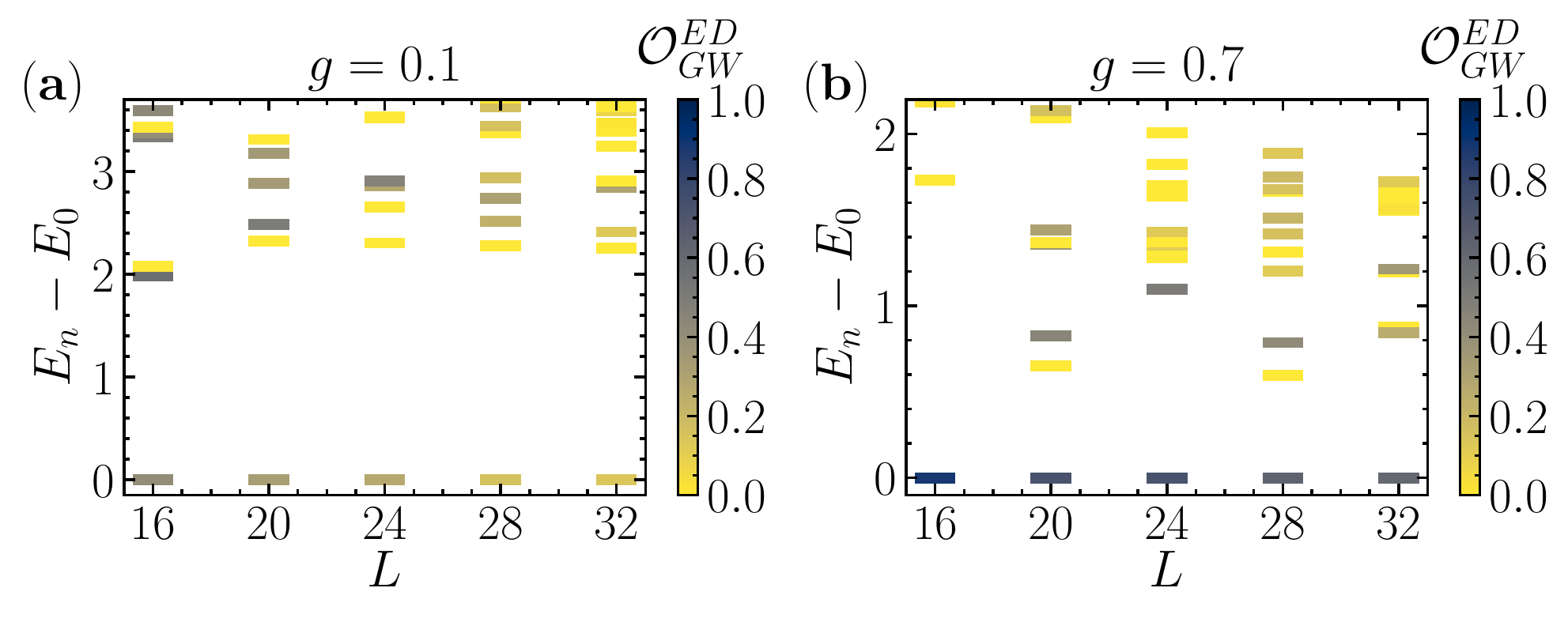}
    \vspace*{-6mm}
    \caption{Excitation spectrum in the momentum sector $k$\,$=$\,$(0,0)$ up to the 10th excited state for (a) $g$\,$=$\,$0.1$ and (b) $g$\,$=$\,$0.7$ at 1/2-density. The markers are colored according to the overlap $\mathcal{O}^{ED}_{GW}$. }
    \label{fig:spectrum}
\end{figure}

In Fig. \ref{fig:spectrum}, we present the excitation spectrum in the momentum sector $k$\,$=$\,$(0,0)$ at $g$\,$=$\,$0.1$ and $g$\,$=$\,$0.7$, along with the overlaps $\mathcal{O}^{ED}_{GW}$ for each eigenstate. Note that, since the two topological states $|\psi_{GW}^{1}\rangle$ and $|\psi_{GW}^{2}\rangle$ lie in the momentum sector $k$\,$=$\,$(0,0)$, only the eigenstates in this sector can have non-zero overlap. In agreement with \cite{Ohler_2022}, we observe no approximate two-fold degeneracy in the ED spectra, which would have been expected in a CSL. Nevertheless, this can be attributed to finite-size effects, which may significantly modify the spectra on small-size clusters. It is therefore possible that one of the low-lying states corresponds to another topological ground state, which becomes degenerate with the true ground state in the thermodynamic limit. To test this hypothesis, it is useful to examine the overlaps of the low-lying levels. If an eigenstate describes the topological ground state of the CSL, it would have a sizable overlap with the wavefunction ansatz. Indeed, at $g$\,$=$\,$0.7$, we observe that the overlap is highest for the ground state, and that there is a low-lying state with a modest overlap. In contrast, at $g$\,$=$\,$0.1$, the overlaps do not exhibit any clear pattern for each system size.

\section{Excitation spectra} \label{sec:spectra}

\begin{figure} 
    \centering
    \begin{overpic}[width=0.45\linewidth]{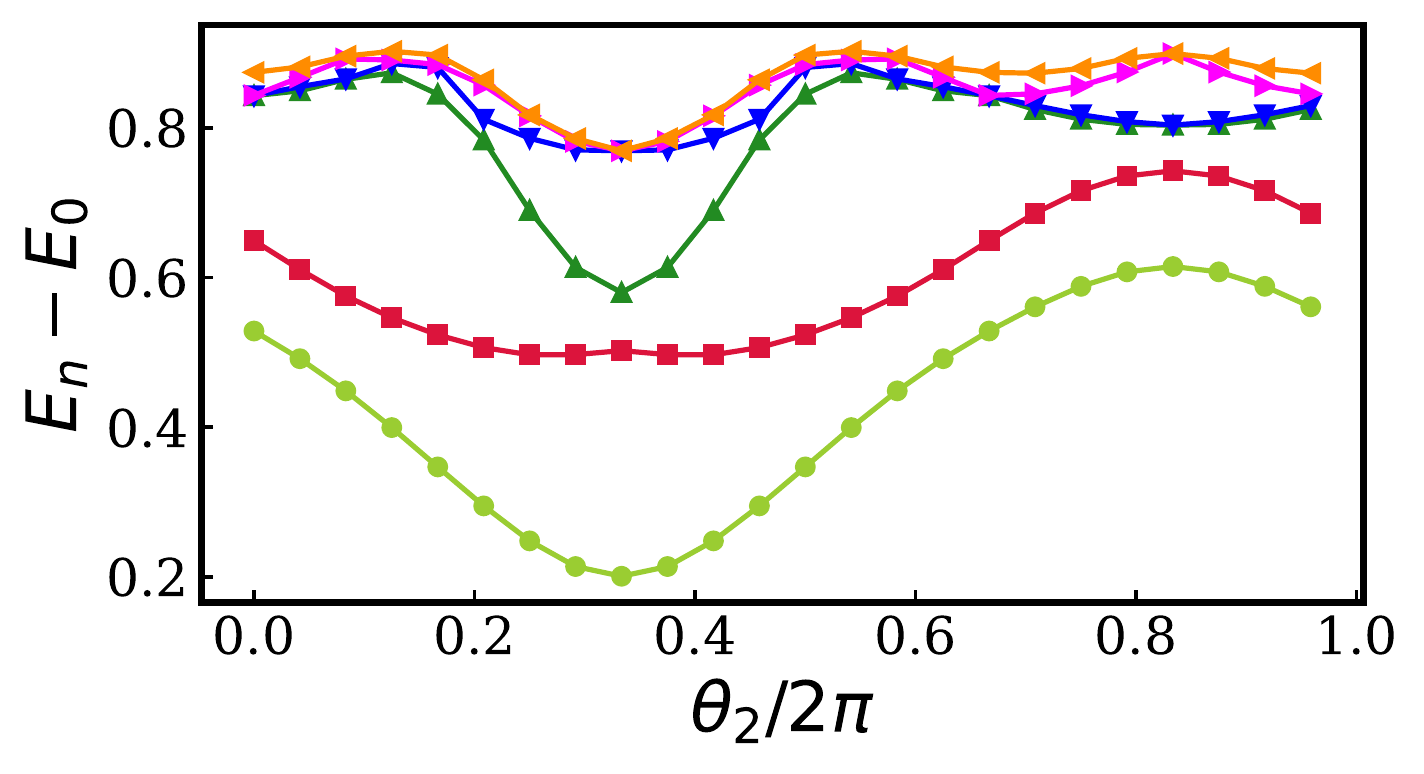}
    \put (-1,60) {{\textbf{(a)}}}
    \end{overpic}
    \begin{overpic}[width=0.45\linewidth]{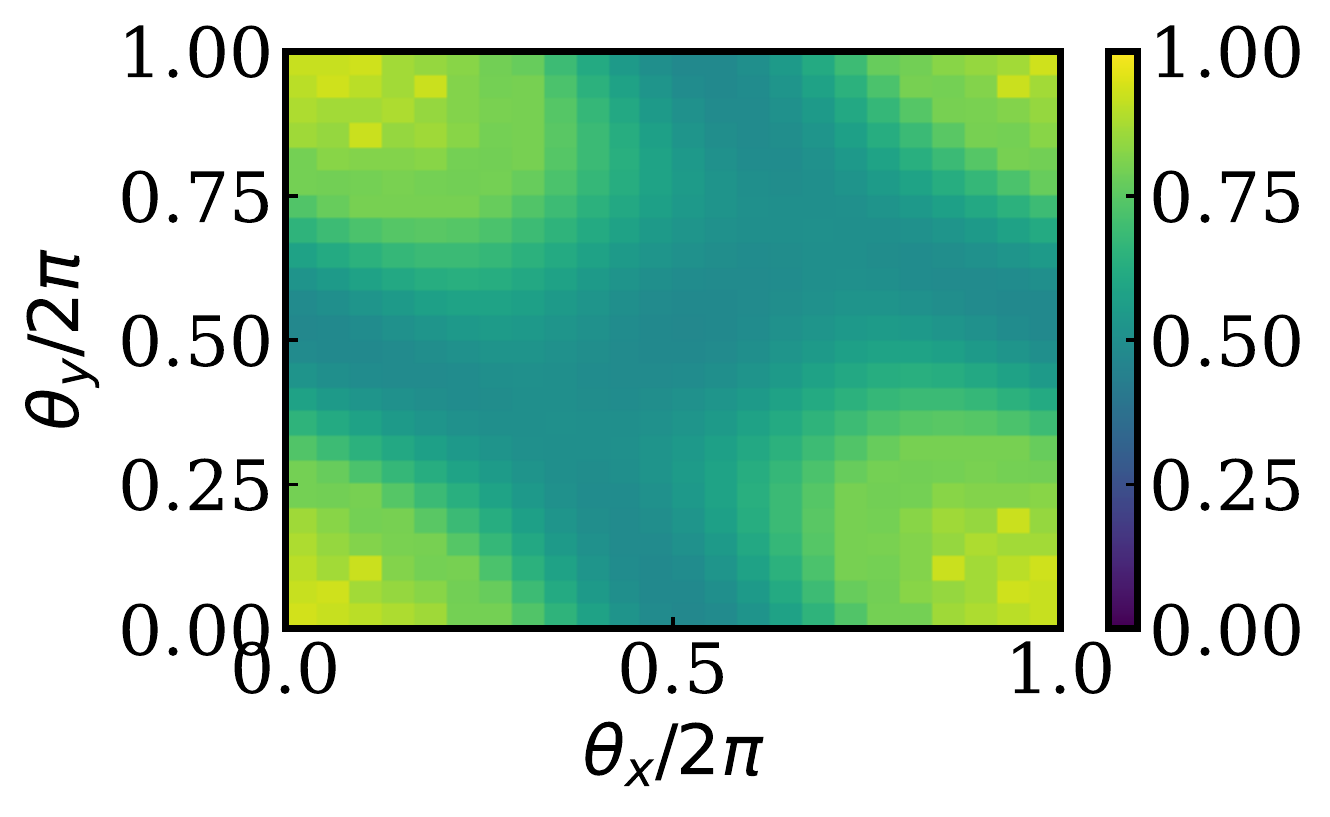}
    \put (-1,64) {{\textbf{(b)}}}
    \end{overpic}
    \begin{overpic}[width=0.45\linewidth]{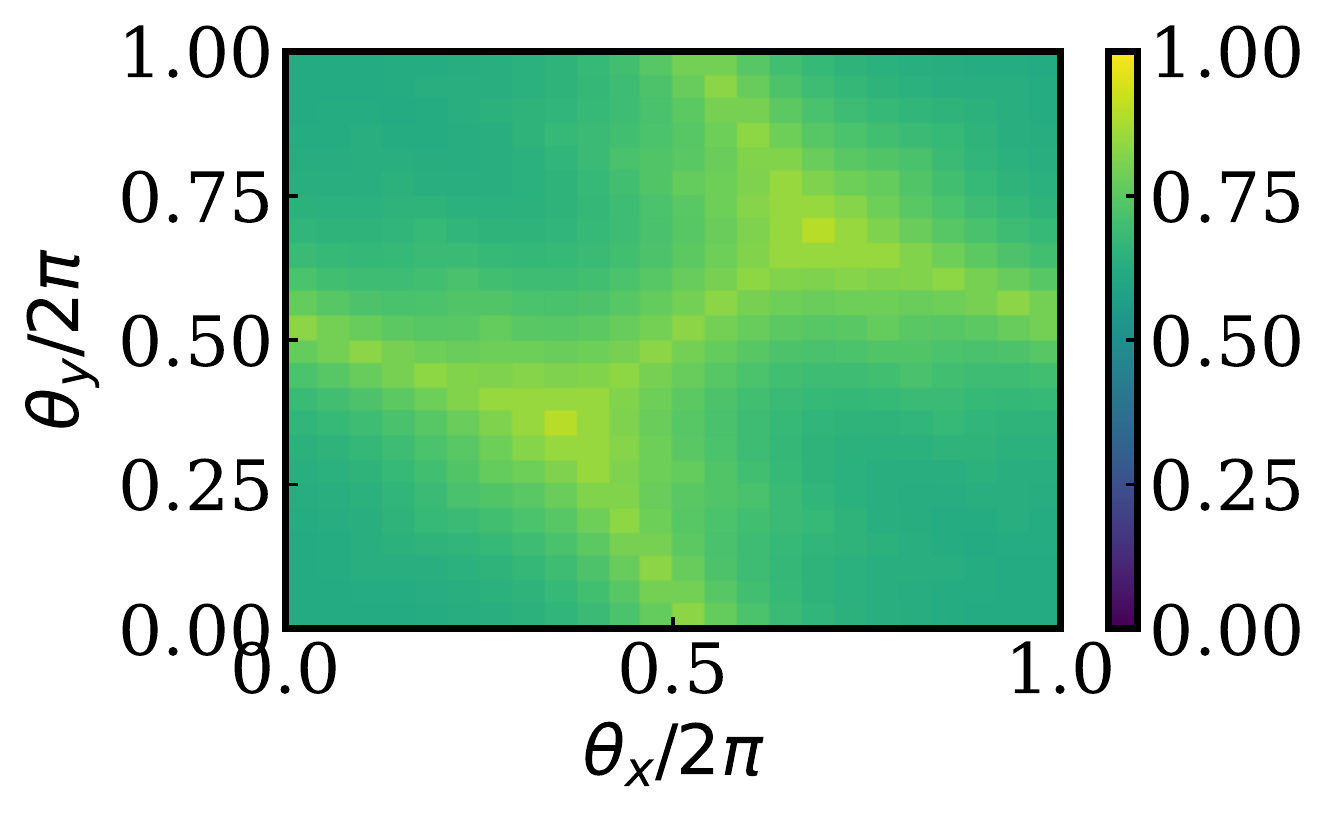}
    \put (-1,64) {{\textbf{(c)}}}
    \end{overpic}
    \begin{overpic}[width=0.45\linewidth]{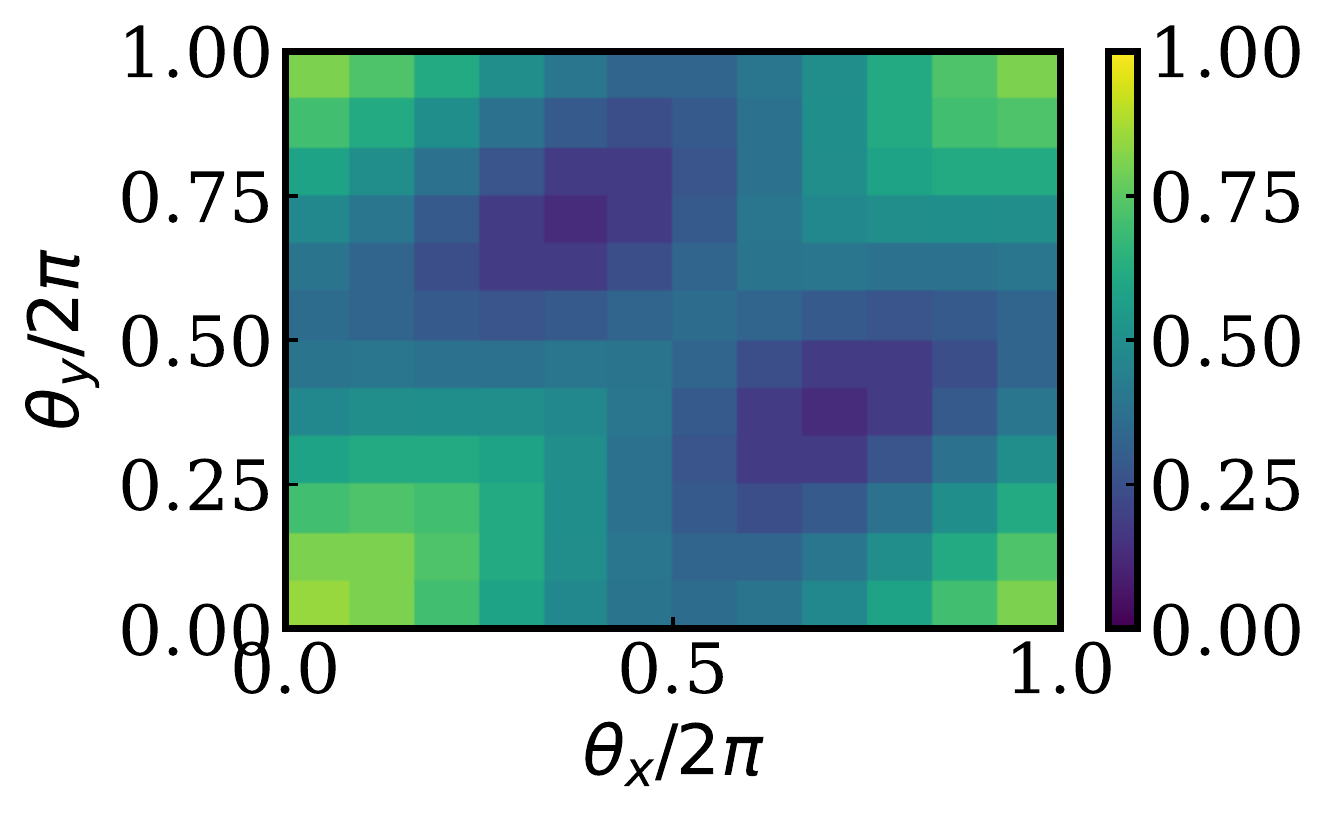}
    \put (-1,64) {{\textbf{(d)}}}
    \end{overpic}
    \caption{(a) Excitation spectrum on the cluster 24a at  $g$\,$=$\,$0.7$ with twist $\theta_1$\,$=$\,$2\pi/3$ and varying $\theta_2$. (b) Gap to the second excited state $E_2-E_0$ and (c) (symmetrized) charge gap $\Delta^c(N)$\,$=$\,$\frac{1}{2}(E(N+1)+E(N-1)-2E(N))$ with $N$\,$=$\,$L/2$ as a function of the twist angles $\{ \theta_1,\theta_2 \}$ on the cluster 24a. (d) Gap to the first excited state $E_1-E_0$ in the $k=(0,0)$ momentum sector on the 32-site cluster as a function of the twist angles $\{ \theta_1,\theta_2 \}$. }
    \label{fig:gaps}
\end{figure}

In Ref. \cite{Ohler_2022}, a DSL was proposed as one of possible scenarios, based on the observation that the gap to the first excited state varies significantly with twist angle when imposing twisted boundary conditions. In light of this, we analyze the excitation gaps as a function of twist angles $\theta_{1,2}$ along the lattice vectors $\vec{a}_{1,2}$. In Fig. \ref{fig:gaps}a, we show the gap to the first excited state obtained with ED on a 24-site cluster, while the gap to the second excited state and the (symmetrized) charge gap are shown in Fig. \ref{fig:gaps}b and \ref{fig:gaps}c, respectively. We observe that while the first gap may become very small at some isolated points in twist space, the second gap and charge gap remain wide open. This contrasts with the expected behavior of a DSL, where all gaps would exhibit a vanishing behavior with respect to twist angles \cite{he2017,hu2019,ferrari2021}. In addition, the transfer matrix spectrum does not show any signatures of Dirac cones (see App. \ref{app:transfer_matrix}). This is consistent with our results that DSL is unstable against time-reversal symmetry breaking perturbations (see Sec. \ref{sec:classification}).

We note that the drastic variation of the first gap with respect to the twist angles becomes even more pronounced at larger sizes, as can be seen in the cluster of $L=32$ sites shown in Fig. \ref{fig:gaps}d. Based on our findings, we are able to offer an interpretation of the curious vanishing of the gap to the first excited state as observed in \cite{Ohler_2022}. Indeed, for a CSL two topological ground states will flow into each other upon inserting flux (see, e.g., \cite{wang2011}).
In the case of CSL at $n$\,$=$\,$1/2$ density, the two ground states live in the same $k$\,$=$\,$(0,0)$ momentum sector. Consequently, the flow between these ground states manifests as an avoided crossing. As a result, the gap appears to vanish at some isolated points, corresponding precisely to the locations where the avoided crossing occurs. 
This behavior is specific to the first gap and not observed in higher gaps or the charge gap, as the CSL is gapped to all excitations. This is consistent with our results on excitation spectra discussed above. 

\section{Topological entanglement entropy} \label{sec:tee}
\begin{figure}
    \centering
    \includegraphics[scale=0.4]{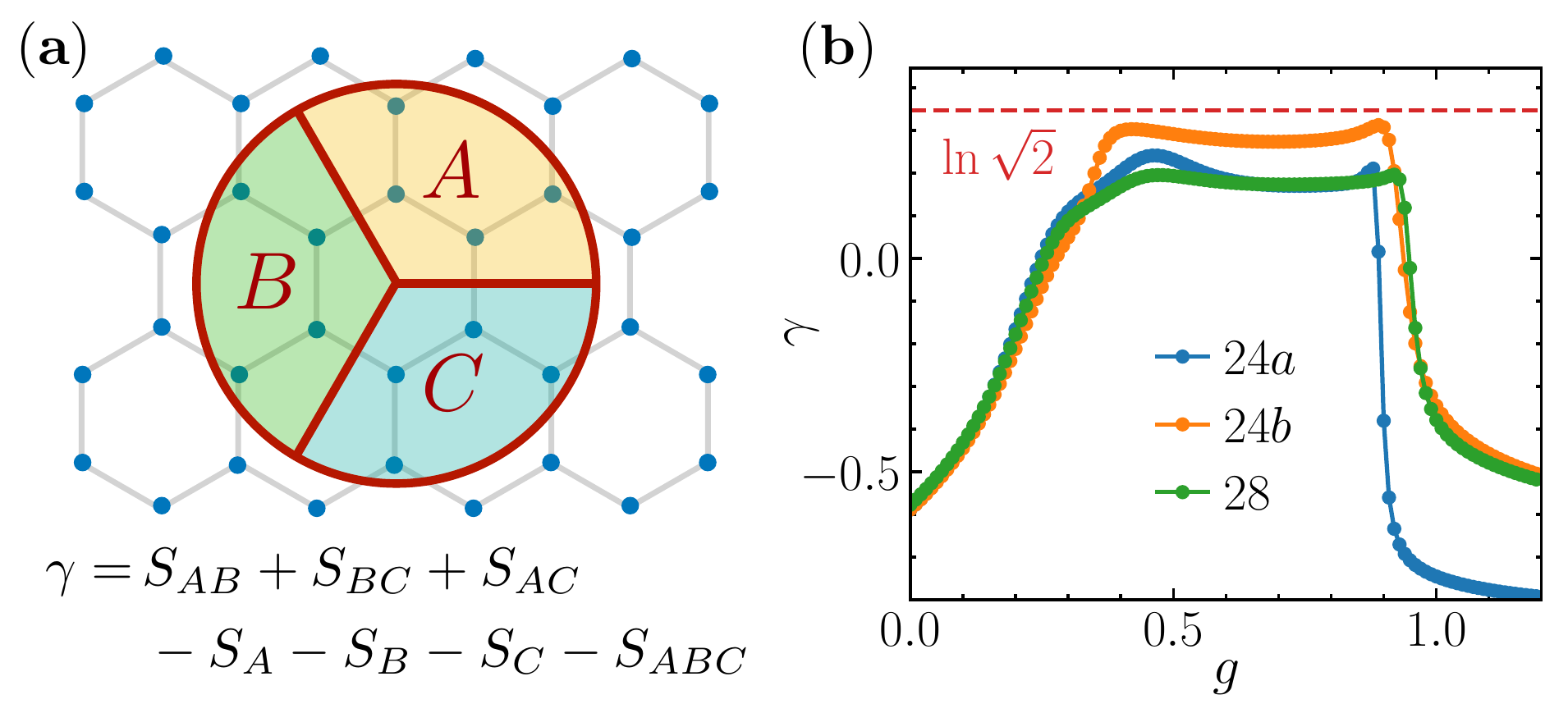}
    \caption{(a) Partitions used for the topological entanglement entropy calculation. (b) Topological entanglement entropy $\gamma$ as a function of $g$ at 1/2-density, obtained from the exact ground state on different periodic clusters (see Fig. \ref{fig:clusters}). The dashed red line is the value of $\gamma$ expected for a $\nu$\,$=$\,$1/2$ FQH state.}
    \label{fig:tee}
\end{figure}

  A topological phase can be characterized by the scaling of the entanglement entropy. The entanglement entropy for a region with perimeter $L$ is known to scale as $S(L)$\,$=$\,$\alpha L$\,$-$\,$\gamma$, where the subleading term $\gamma$ is a universal constant called the topological entanglement entropy which characterizes the topological order in a ground state wavefunction \cite{kitaev2006,levin2006}. Here, we employ the Kitaev-Preskill scheme \cite{kitaev2006} to compute $\gamma$:
  \begin{equation}
        \gamma = S_{AB} + S_{BC} + S_{AC} - S_{A} - S_{B} - S_{C} + S_{ABC},
    \end{equation}
  where the partitioning is depicted in Fig. \ref{fig:tee}a. In Fig. \ref{fig:tee}b, we show the behavior of $\gamma$ in the exact ground state obtained with ED for $n$\,$=$\,$1/2$ density. In the CSL phase, even in small systems, the computed $\gamma$ is finite, and close to the expected value $\gamma$\,$=$\,$\frac{1}{2}\log(2)$ for a $\nu$\,$=$\,$1/2$ FQH state~\cite{footnote6}.

\section{Chiral currents} \label{sec:currents}
Quantum Hall states can be identified through the pattern of the currents in the system. With a finite gap in the bulk and gapless edge excitations on the boundary, it is expected that the currents are large at the edges and vanish in the bulk. Furthermore, the current can be readily measured in experiments, making it a convenient tool for diagnosing the phase in experimental setups. The NN and NNN currents can be derived using the continuity equation, resulting in 
\begin{align} 
& \mathcal{J}_{\text{nn}}=i\langle b^\dagger_j b_i - b_j b^\dagger_i \rangle, \nonumber \\
& \mathcal{J}_{\text{nnn}}=2 g i e^{\pm 2\pi/3 i}\langle (1-n_{i j}) (b^\dagger_j b_i - b_j b^\dagger_i) \rangle,
\label{eq:current}
\end{align}
where $i$ and $j$ are nearest-neighbors or next to nearest-neighbors respectively.

\begin{figure}
    \centering
\includegraphics[width=0.7\linewidth]{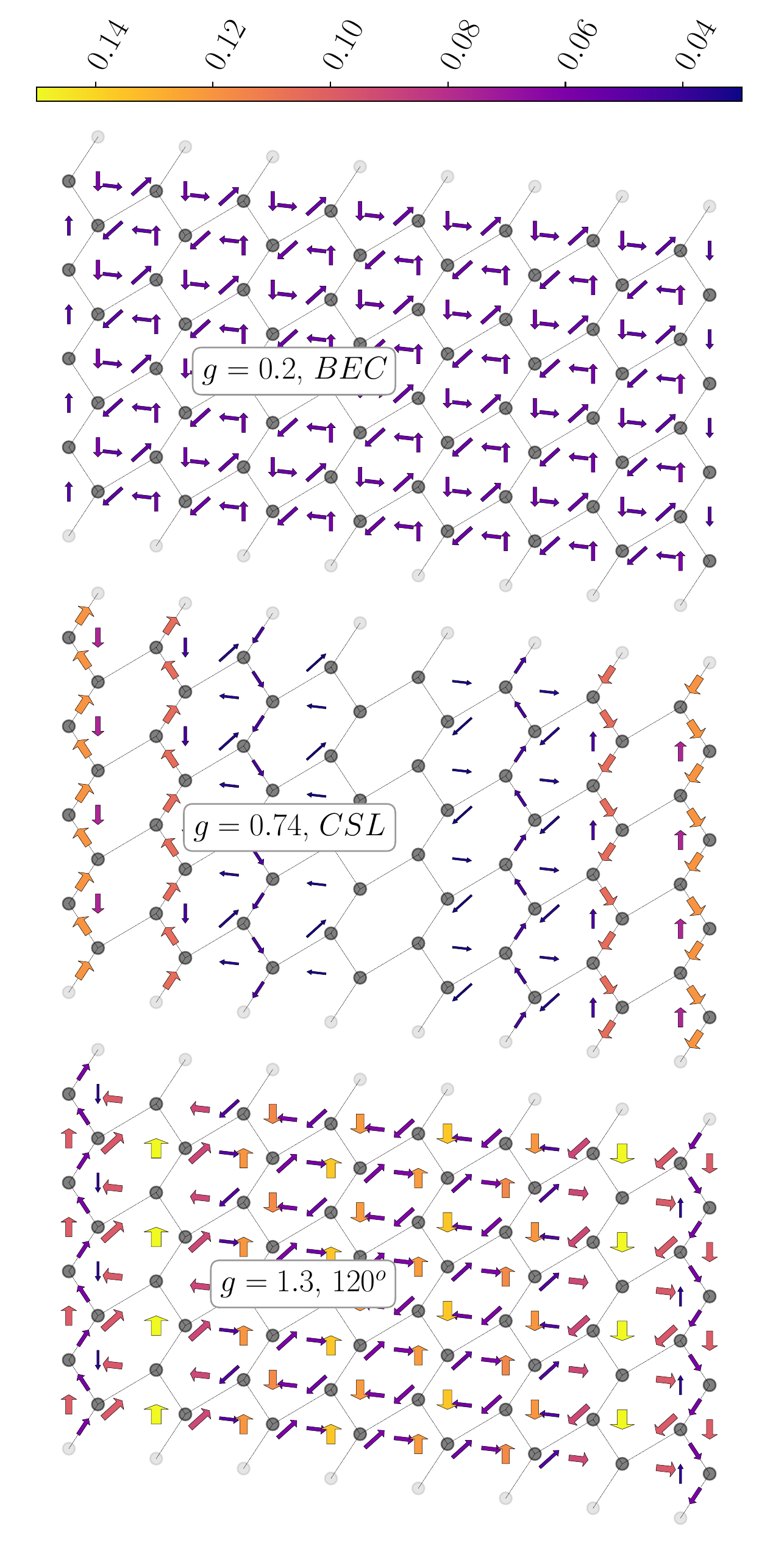}
    \caption{Full current profiles at $n=1/2$ filling for the cylinders with geometry I and  with periodic width $L_{\text{PBC}}= 4$ and length $L_{\text{OBC}} = 8$ for three representative values of $g$ inside three phases, namely BEC (top), CSL (middle), and the $120^{\circ}$ (bottom) phases. The widths of the arrows are proportional to the magnitudes of the current, while their direction indicates the directions of respective currents.
    }
    \label{fig:current}
\end{figure}

We perform DMRG simulations on a finite cylinder to compute the  edge currents as a function of $g$, as shown in Fig. \ref{fig:schematic}c~\cite{footnote4}. It can be seen that the NN edge currents, computed across two rungs in one of the edges, in the intermediate phase are significantly larger compared to those in the neighboring phases. The transition points are in good agreement with those found in \cite{Ohler_2022}. Furthermore, we show the full current profile in the intermediate phase ($g$\,$=$\,$0.74$) in Fig. \ref{fig:schematic}d. It is clear that, in the CSL phase, large NN currents are only observed at the edges, while they vanish in the bulk. The full current profile is shown for each phases in Fig. ~\ref{fig:current}. Clearly, substantial counter-propagating NN edge currents manifests only in the 
CSL phase that vanish in the bulk. On the other hand, most dominant currents in the other phases are of the NNN nature that originate from the NNN term in the Hamiltonian. 


\begin{figure*} [ht]
    \includegraphics[scale=0.47]{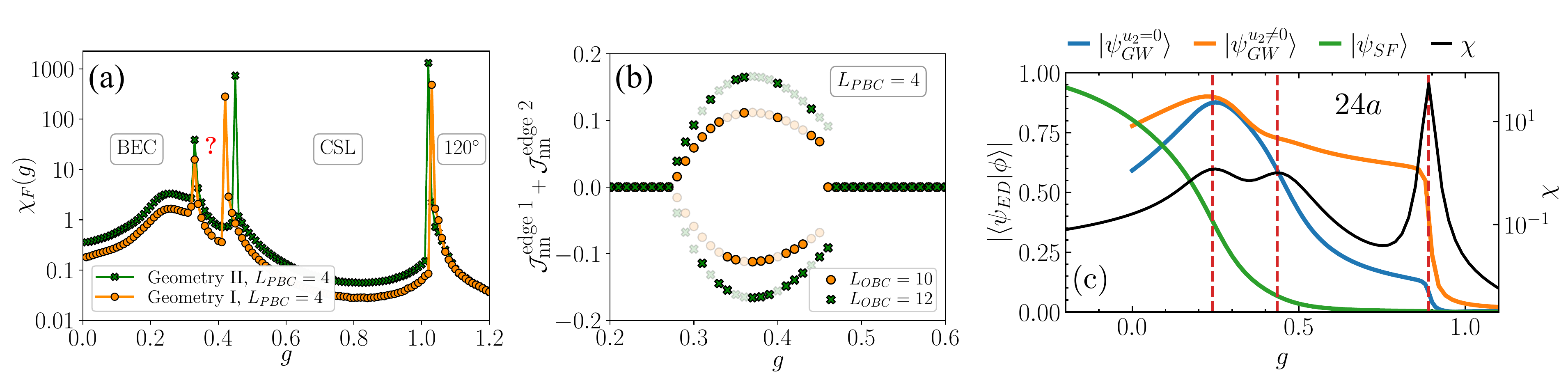}
    \caption{(a) The fidelity susceptibility $\chi_{F}(g)$ as a function of $g$ for infinite cylinders with two different geometries and  with periodic width $L_{\text{PBC}} = 4$ in the $n=1/2$ density regime. The $\chi_{F}(g)$ shows the existence of a new phase sandwiched between the BEC and the CSL phases. (b) The total nearest-neighbor edge current $\mathcal{J}^{\text{edge 1}}_{\text{nn}} + \mathcal{J}^{\text{edge 2}}_{\text{nn}}$ computed for finite cylinders of geometry I and with periodic width $L_{\text{PBC}} = 4$ for $n=1/2$ density.
    In the new intermediate phase, reflection~$\times$~time-reversal symmetry gets spontaneously broken, and the total edge current becomes finite (positive or negative). Since, the DMRG simulations breaks this $\mathbb{Z}_2$ symmetry somewhat randomly, there is a arbitrariness in the sign of the edge current. That is why 
    transparent symbols are added that represent the currents with opposite signs of the actual DMRG data. 
    (c) Fidelity susceptibility and overlaps of the exact ground state with various ansatz on the cluster 24a for $n=1/2$. Orange line: overlap with ansatz no. 1 optimized at each point. Blue line: Overlap with DSL wavefunction with uniform spin-dependent NN hopping $\tau^0$. Green line: Overlap with superfluid ansatz,  defined as the equal-weight superposition of all states at fixed density. The emergence of the additional phase can already be observed, between the two nearby peaks of fidelity susceptibility at $g=0.24$ and $g=0.43$, which are denoted by the vertical red lines. The overlap with DSL wavefunction peaks close to the transition between the BEC phase and the additional phase.  }
    \label{fig:fidelity}
\end{figure*}
\section{Another intermediate phase} \label{sec:new_phase}

Notice that Fig. \ref{fig:schematic}c also exhibits signatures of another intermediate phase with non-zero edge currents for $0.25$\,$\lesssim$\,$g$\,$\lesssim$\,$0.4$. While its full characterization is beyond the scope of this work, we report here our preliminary findings on this possible new phase.

To detect transition points and obtain the phase diagram as a function of $g$, we compute the fidelity-susceptibility $\chi_F$ defined as
\begin{equation}
    \chi_F(g)= \frac{2}{L} \lim_{\delta g \rightarrow 0} \frac{-\ln F(g,\delta g)}{(\delta g)^2}
\end{equation}
where the fidelity $F$ is defined as $F(g,\delta  g)=|\langle \psi(g) | \psi(g+\delta g) \rangle|$. 
The fidelity susceptibility  $\chi_F$ is known to be a good indicator for quantum phase transition, whose critical point can be derived via finite-size scaling techniques \cite{Gu2009}.
In case of iDMRG simulations, however, we have
\begin{equation}
    F(g,\delta  g) = \lim_{L \rightarrow \infty} |\eta|^{L},
\end{equation}
where $\eta$ is the dominant eigenvalue of the transfer matrix constructed from the iMPS ansatze of $\ket{\psi(g)}$ and  $\ket{\psi(g+\delta g)}$. Since, for two normalized iMPS $\ket{\psi(g)}$ and  $\ket{\psi(g+\delta g)}$, $|\eta| < 1$ for $\delta g \neq 0$, we get $F(g,\delta  g) \rightarrow 0$ as $L \rightarrow \infty$. The fidelity-susceptibility can be expressed as
\begin{equation}
    \chi_{F} (g) = - 2 \ln|\eta|,
\end{equation}
which remains finite.

We show the fidelity-susceptibility obtained using iDMRG for two different types of cylinders in Fig. \ref{fig:fidelity}(a). The emergence of the CSL phase can be identified in the parameter range $0.4 \lesssim g \lesssim 1$. Interestingly, there appears to be an additional phase emerging between the BEC phase and the CSL. Since, the system is invariant under the joint operation of time-reversal and reflection, total nearest-neighbor edge current $\mathcal{J}^{\text{edge 1}}_{\text{nn}} + \mathcal{J}^{\text{edge 2}}_{\text{nn}}$, odd under this joint operation, must vanish for the ground states that respects reflection~$\times$~time-reversal symmetry. By performing finite DMRG on finite-cylinders, we find that this reflection~$\times$~time-reversal symmetry gets spontaneously broken in this parameter regime, and the total nearest-neighbor edge current $\mathcal{J}^{\text{edge 1}}_{\text{nn}} + \mathcal{J}^{\text{edge 2}}_{\text{nn}}$ attains a finite (positive or negative) value (see Fig.~\ref{fig:fidelity}(b)). The profile of the total edge current in Fig.~\ref{fig:fidelity}(b) clearly indicates two-fold degenerate ground-state manifold in this new unknown phase.

We note that the emergence of this additional phase can already be seen in small size clusters, as observed from ED calculations. In Fig. \ref{fig:fidelity}(c), we show the fidelity susceptibility as well as the overlap of the exact ground state with various ansatz on the cluster 24a. We find that the additional phase also have a large overlap with ansatz no. 1, where now we optimize the ansatz at each point (orange line). We find that the optimal ansatz in this phase is in the vicinity of the DSL wavefunction. Indeed, the overlap with the DSL wavefunction (blue line), which is obtained by setting the NNN interactions to zero, is remarkably close with the optimal overlap. One possible scenario to explain this observation is that the DSL wavefunction describes the critical wavefunction at the transition from the BEC phase to the unknown phase. A similar scenario was put forward in \cite{Wietek_2017}. Finally, the optimal overlap becomes smaller in the BEC phase, as expected. Instead, in this regime, large overlaps are obtained with the superfluid ansatz, which is defined as the equal-weight superposition of all states at fixed density \cite{Wen2013}.

\section{Conclusions} \label{sec:concl}
In this work, we systematically classify CSLs on the honeycomb lattice relevant to Rydberg atom experiments using the PSG analysis. We show that the CSL wavefunctions constructed from the Gutzwiller-projected parton wavefunctions are able to capture the intermediate disordered phase in chiral Rydberg atom arrays. In particular, our results resolve the previously unclear nature of the intermediate phase found in Ref.~\cite{Ohler_2022}. In the context of Rydberg atom experiments, our work provides a general framework which can be utilized to search for CSLs in other lattice models. Given the fast experimantal progress in the field, it would be interesting to extend our approach to other lattices, which are immediately available in tweezer arrays~\cite{browaeys_2020,barredo2018synthetic}. 

 \begin{acknowledgments}
We thank F. Becca, M. Fleischhauer, and S. Ohler for insightful discussions. M.D. and P.S.T. are grateful to Y. Iqbal for explanations and illuminating clarifications on the parton construction.
The work of M.D. and P.S.T. was partly supported by the ERC under grant number 758329 (AGEnTh), by the MIUR Programme FARE (MEPH), and by the European Union's Horizon 2020 research and innovation programme under grant agreement No 817482 (Pasquans). P.S.T. acknowledges support from the Simons Foundation through Award 284558FY19 to the ICTP.
G.G. acknowledges support from the European Union’s Horizon Europe program under the
Marie Sklodowska Curie Action TOPORYD (Grant No. 101106005), from the Deutsche Forschungsgemeinschaft (DFG, German Research Foundation) under Germany’s Excellence Strategy – EXC2111 – 390814868 and from the ERC grant QSIMCORR, ERC-2018-COG, No.~771891. M. D. further acknowledges support from QUANTERA DYNAMITE PCI2022-132919.
T.C. acknowledges the support of PL-Grid Infrastructure for providing high-performance computing facility for a part of the numerical simulations reported here.

\end{acknowledgments}

\appendix

\section{Adiabatic elimination and truncation of the dipolar interactions}
\label{app:approx}
\begin{figure} [h]
    \centering
    \includegraphics[scale=0.44]{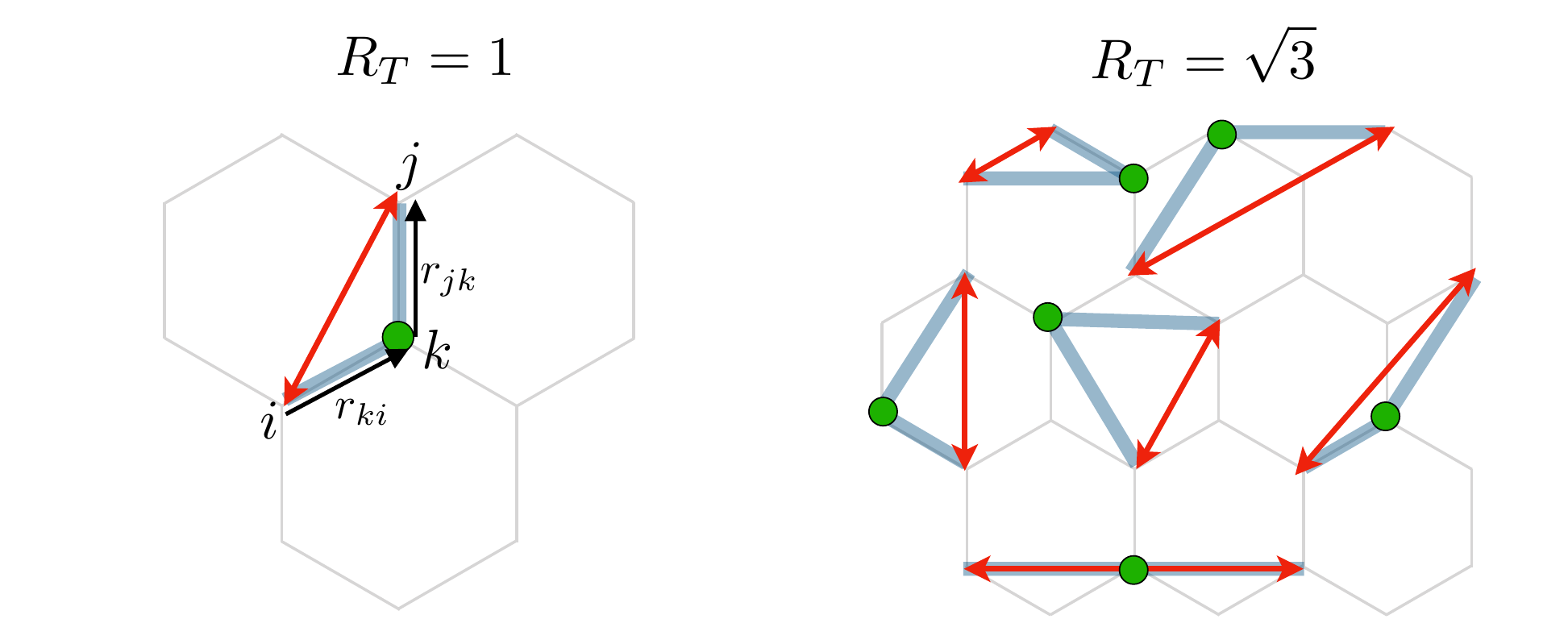}
    \caption{Schematic representation of the virtual processes generating the density-assisted chiral hopping in the limit $\mu \to \infty$ when $R_T = 1$ (left) and $R_T = \sqrt{3}$ (right).}
     \label{fig:1}
\end{figure}

\noindent
In the main text we introduced the following Hamiltonian
\begin{align} 
    \nonumber 
    H_0 = &\sum_{i \neq j} \begin{pmatrix} a_i \\ b_i \end{pmatrix}^\dagger \begin{pmatrix}
         -t^a_{ij} && w_{ij} e^{-i 2\phi_{ij}} \\ w_{ij} e^{i 2\phi_{ij}} && -t^b_{ij} \end{pmatrix} \begin{pmatrix} a_j \\ b_j \end{pmatrix} \\
         &+ \frac{\mu}{2} \sum_i (n^a_i- n^b_i). 
         \end{align}
where $i$ labels the sites of a honeycomb lattice, $t_{ij} = t/d_{ij}^3$, $w_{ij} = w/d_{ij}^3$, and $d_{ij}$ is the distance between sites $i$ and $j$. The hopping phase $\phi_{ij}$ is the angle between the position vectors of sites $i$ and $j$. The operators $a^\dagger_i$ and $b^\dagger_i$ create hard-core bosons ($(a^\dagger_i)^2 = (b^\dagger_i)^2 = 0 $) on site $i$, subject to the constraint $a^\dagger_i b^\dagger_i = 0$ (at most one particle per site). \\
We argued that when $\mu \gg t_{ij},w$ one can adiabatically eliminate the $a$ particles, and we considered an effective model where the particle created by $a^\dagger$ is integrated out. Thus, the effective model only includes the hard-core bosonic particle created by $b^\dagger$, while $a$ particles only appear in virtual processes. When the hopping amplitudes are truncated to nearest-neighbor distance $R_T = 1$ (in units of one lattice spacing) the effective Hamiltonian reads 
\begin{align}
\nonumber
        H = &- \sum_{\langle ij \rangle} b^\dagger_j b_i - 2 g  \sum_{\langle \langle ij \rangle \rangle} b^\dagger_j b_i e^{ s_{ikj} 2\pi i/3} (1-n_k) + \text{h.c.} \\
        &+ 4 g  \sum_{\langle ij \rangle} n_i n_j, 
\end{align}
where $g = w^2/4 t$, the site $k$ is the only possible site between the next-nearest-neighbor sites $i$ and $j$ that can be reached with two virtual nearest neighbor hoppings, and $s_{ikj} = \mathrm{sign}( r_{jk} \times r_{ki} )$ (see Fig~\ref{fig:1}). The density-density interaction term $n_i n_j$ comes from the virtual process of the form $(b_i a^\dagger_k) (a_j b^\dagger_i) = b_i n^a_j b^\dagger_i$ upon dropping a constant term in the effective Hamiltonian, and its amplitude is \emph{twice} the amplitude of the density-assisted chiral hopping due to the fact that it gets contributions from two virtual processes, starting from sites $i$ and $j$.\\
The effective Hamiltonian Eq. \eqref{eq:model} lives in a reduced Hilbert space and it is thus more amenable to numerical calculations. However, its fairly simple form relies on the truncation of the hopping coefficients $t_{ij}$ and $w_{ij}$ at nearest-neighbor distance. Extending the range of the real hopping $R_T$ up to next-nearest-neighbor ($R_T = \sqrt{3}$), leads to $6$ extra chiral hopping terms which are depicted in Fig.~\ref{fig:1}, as well as an additional density-density interaction term between next-nearest-neighbor sites.\\

In this section we discuss the effect of finite $\mu$ and of larger truncation distances $R_T$.
We do so by performing exact diagonalization for the ground state of the model Eq.~\eqref{eq:fullModel} both in the limit $\mu \to \infty$ and at finite $\mu$, when $R_T = 1$ and $R_T = \sqrt{3} $. We refer to ``truncated model" in connection to hopping truncation, and to ``effective model" in connection to truncation of the Hilbert space following elimination of the $a$ particles. For simplicity we focus on the filling sector where the total number of particles is $N/4$, where $N$ is the number of sites. In fact, the following analysis is aimed at probing the validity of the adiabatic elimination at different hopping truncations $R_T$, and does not aim at probing the stability of a CSL phase at finite $\mu$ and $R_T > 1$ in a generic filling sector. 

\begin{figure} [t]
    \centering
    \includegraphics[scale=0.35]{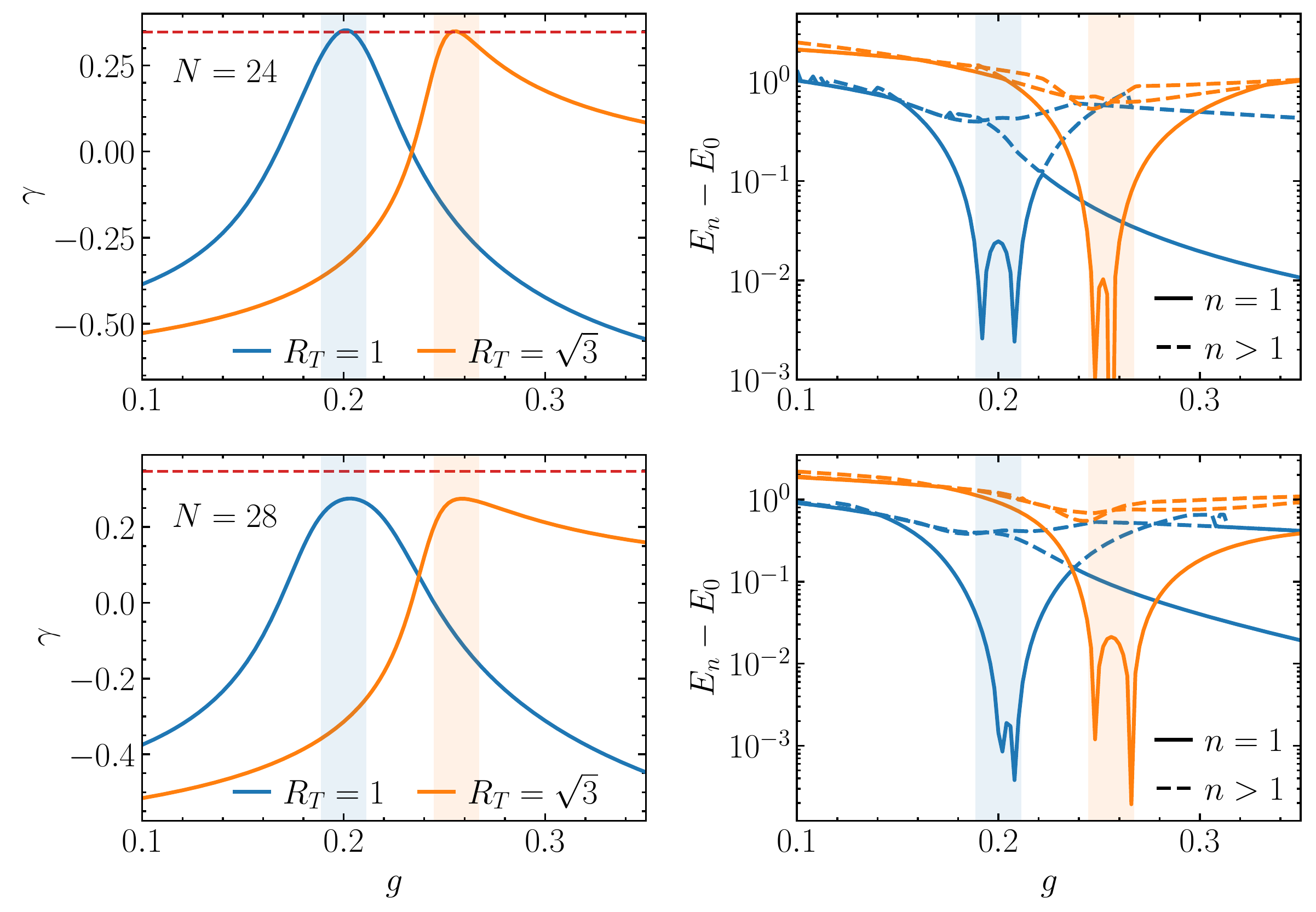}
    \caption{Ground state topological entanglement entropy $\gamma$ (left) and low-lying gaps of the effective Hamiltonian (right) when $\mu \to \infty$ and for $R_T = a$ (blue) and $R_T = \sqrt{3} a$ (orange), for $N=24$ (top) and $N=28$ (bottom). The dashed red horizontal line on the left panels denotes the value $\gamma = \log \sqrt{2}$ expected for a chiral spin liquid ground state. }
     \label{fig:2}
\end{figure}

We now provide numerical evidence from exact diagonalization on periodic clusters of $24$ and $28$ sites that the $\mu \to \infty$ effective model hosts a chiral spin liquid phase for both $R_T = 1$ and $R_T = \sqrt{3}$. Fig.~\ref{fig:2} shows the topological entanglement entropy $\gamma$ extracted from the partitions depicted in Fig. 4 of the main text, and the lowest eigenvalues of the effective Hamiltonian. Whereas the former exhibits a bump approaching the value $\log \sqrt{2}$, the latter shows an approximate two-fold degeneracy of the ground state in the same narrow parameter range, pointing at the emergence of a chiral spin liquid (CSL) regime. We note that the extending $R_T$ from $1$ to $\sqrt{3}$ has two effects: it shifts to the right the CSL phase and it changes the nature of the large-$g$ phase, as can be inferred from the different value of $\gamma$ and structure of the energy levels.\\

\begin{figure} [h]
    \centering
    \includegraphics[scale=0.45]{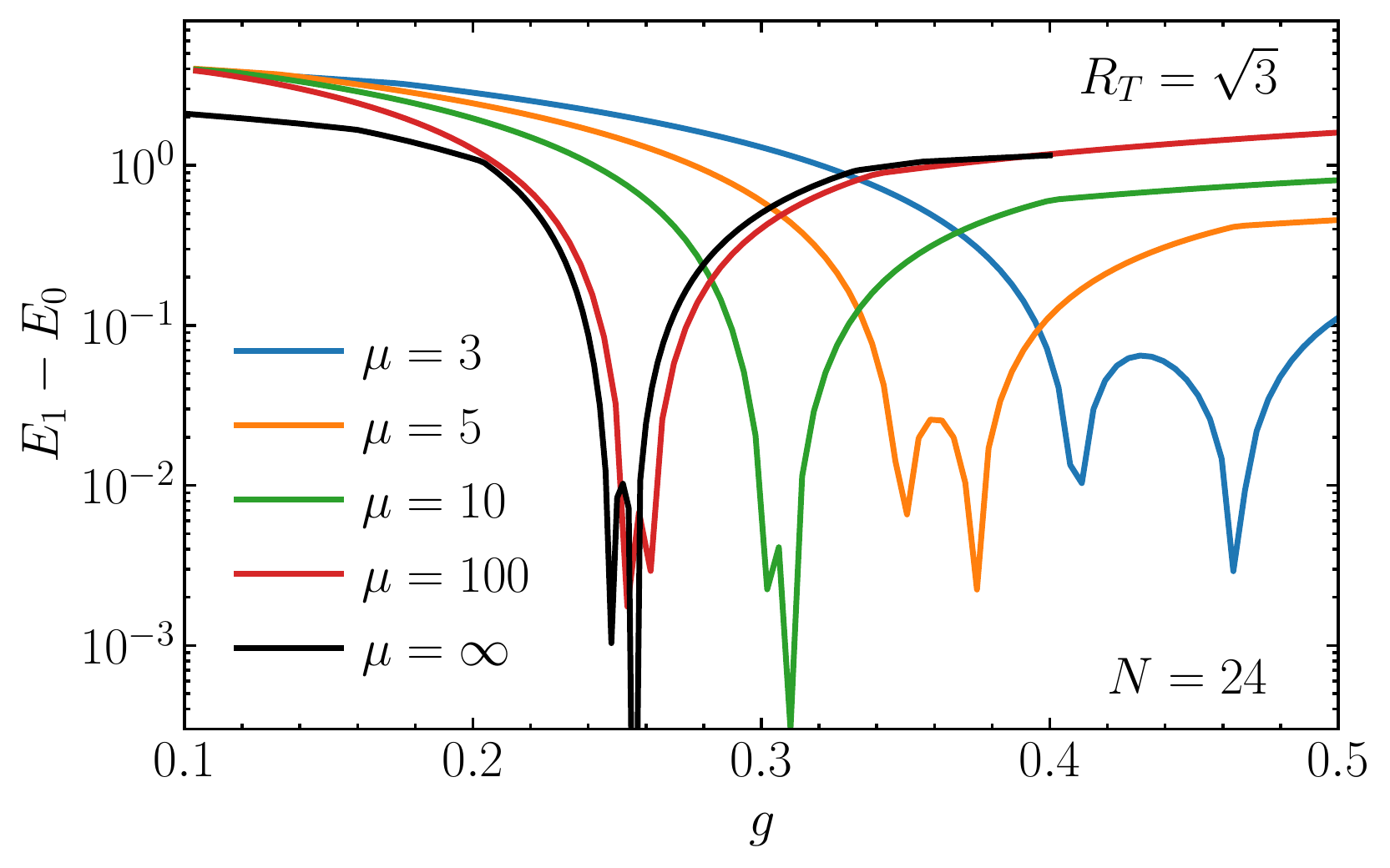}
    \caption{Lowest energy gap for several values of $\mu$ on a periodic cluster with $N=24$ for the full model Eq.~\eqref{eq:fullModel}. }
     \label{fig:3}
\end{figure}

We conclude by studying the effect of finite $\mu$ on the ground state of the Hamiltonian \eqref{eq:fullModel} with $R_T = \sqrt{3}$. In Fig.~\eqref{fig:3} we plot the lowest energy gap for several values of $\mu$ ranging from $3$ to $100$. We observe convergence (uniformly in $g$) to the $\mu = \infty$ curve obtained from the effective model with chiral hopping terms depicted on the right of Fig.~\eqref{fig:1}. Despite the lowest energy gap is not enought to unequivocally identify a CSL phase at finite $\mu$, its functional dependence on $g$ assumes more relevance when compared to the data presented in Fig.~\eqref{fig:2} and suggests a widening of the CSL regime with decreasing $\mu$ (that, we note, is not monotonous).\\

\section{More on PSG classification} \label{app:psg}
 The symmetry group on the honeycomb lattice that we are interested in is generated by translations along $x$ and $y$, reflection accompanied by time-reversal $\mathbf{T}\sigma$, and $\pi/3$ rotation centered on a hexagonal plaquette $R$. Each algebraic PSG class is characterized by the SU(2) representation of each symmetry generator: $g_x(x,y,s), g_y(x,y,s)$,$ g_\sigma(x,y,s)$, $g_R(x,y,s)$, respectively. This can be further simplified by working in a specific gauge. Here, we choose a gauge defined as~\cite{Lu_2011}
\begin{eqnarray}
    &&g_x(x,y,s) = \tau^0  \\
    &&g_y(x,y,s)  = \epsilon^x \tau^0 \\
    &&g_{\sigma}(x,y,s) =  \epsilon^{x+y(y+1)/2} g_{\sigma}(s) \\
    &&g_R(x,y,s)  =  \epsilon^{xy+x(x-1)/2} g_R(s),
\end{eqnarray}
where $\epsilon=\pm 1$, with $\epsilon=-1$ indicates that the unit-cell is doubled in the spinon space. Within this gauge choice, each PSG is characterized by the representations of reflection, $g_\sigma(A,B)$, and $\pi /3$ rotation, $g_R(A,B)$, for each sublattice $A$ and $B$. The representation matrices satisfy the equations (for a detailed derivation, see \cite{Lu_2011})
\begin{eqnarray}
    &&g_{\sigma}(A) g_{\sigma}(B)  = g_{\sigma}(B) g_{\sigma}(A) = \epsilon_{\sigma} \tau^0 \\
    &&(g_{\sigma}(A) g_{R}(B))^2  = (g_{\sigma}(B) g_{R}(A))^2 = \epsilon_{\sigma R} \tau^0 \\
    &&(g_{R}(A) g_{R}(B))^3  = (g_{R}(B) g_{R}(A))^3 = \epsilon \epsilon_{R} \tau^0 .
\end{eqnarray}
We find 24 different classes of algebraic PSG, which are listed in table \ref{table:psg}. 

\begin{table}
\centering
    \begin{tabular}{|c|c c|c c c|} 
    
    \hline
        No.   &  $g_\sigma(A/B) $ & $g_R(A/B)$ &  $\epsilon_\sigma$ & $ \epsilon \epsilon_R$ & $\epsilon_{\sigma R}$ \\
        \hline
         $1$ & $\tau^0$/$\tau^0$ & $\tau^0/\tau^0$ &  + &  + & +\\
         \hline
         $2$ & $-\tau^0$/$-\tau^0$ & $\tau^0/-\tau^0$ &  + &  - & +\\
         \hline
         $3$ & $\tau^0$/$-\tau^0$ & $\tau^0/\tau^0$ &  - &  + & +\\
         \hline
         $4$ & $-\tau^0$/$\tau^0$ & $\tau^0/-\tau^0$ &  - &  - & +\\
         \hline
         $5$ & $-i\tau^2$/$i\tau^0$ & $\tau^0/\tau^0$ &  + &  + & -\\
         \hline
         $6$ & $-i\tau^2$/$i\tau^0$ & $\tau^0/-\tau^0$ &  + &  - & -\\
         \hline
         $7$ & $i\tau^2$/$i\tau^0$ & $\tau^0/\tau^0$ &  - &  + & -\\
         \hline
         $8$ & $i\tau^2$/$i\tau^0$ & $\tau^0/-\tau^0$ &  - &  - & -\\
         \hline
         $9$ & $-i\tau^2$/$i\tau^0$ & $\tau^0/a$ &  + &  + & -\\
         \hline
         $10$ & $-i\tau^2$/$i\tau^0$ & $\tau^0/-a$ &  + &  - & -\\
         \hline
         $11$ & $i\tau^2$/$i\tau^0$ & $\tau^0/a$ &  - &  + & -\\
         \hline
         $12$ & $i\tau^2$/$i\tau^0$ & $\tau^0/-a$ &  - &  - & -\\

    \hline
    
    \end{tabular}
\caption{PSG representations of point-group symmetries on the honeycomb lattice.Taking into  account the two possible signs $\epsilon=\pm 1$ gives 24 distint algebraic PSG's in total.}
    \label{table:psg}
\end{table}

For each algebraic PSG, we now determine the mean-field amplitudes $u_{ij}$ allowed by symmetry up to NNN links. The consistency conditions for $u_1$ and $u_2$ are
\begin{equation}
\begin{aligned}
    u_1^\dagger = -(-1)^\chi g_\sigma(A) u_1 g_\sigma(B)^\dagger \\
    u_1^\dagger =  g_R(A) g_R(B) g_R(A)  u_1 g_R(B)^\dagger g_R(A)^\dagger g_R(B)^\dagger \\
    u_2^ =  -(-1)^\chi g_R(B) g_\sigma(A)  u_2 g_\sigma(B)^\dagger g_R(A)^\dagger, \\
\end{aligned}
\end{equation}
where $\chi=0$ for singlet terms and $\chi=1$ for triplet terms. The 6 solutions are shown in the main text. Each $u_{ij}$ can be propagated to the entire lattice using rotations, which act as
\begin{equation}
u_{ij} = g_R(i) u_{R^{-1}(i)R^{-1}(j)} g_R(j)^\dagger,
\end{equation}
followed by translations, which act similarly with $g_{x,y}$.

\section{Overlaps} \label{app:overlaps}
\begin{figure}
    \centering
    \begin{overpic}[width=0.45\linewidth]{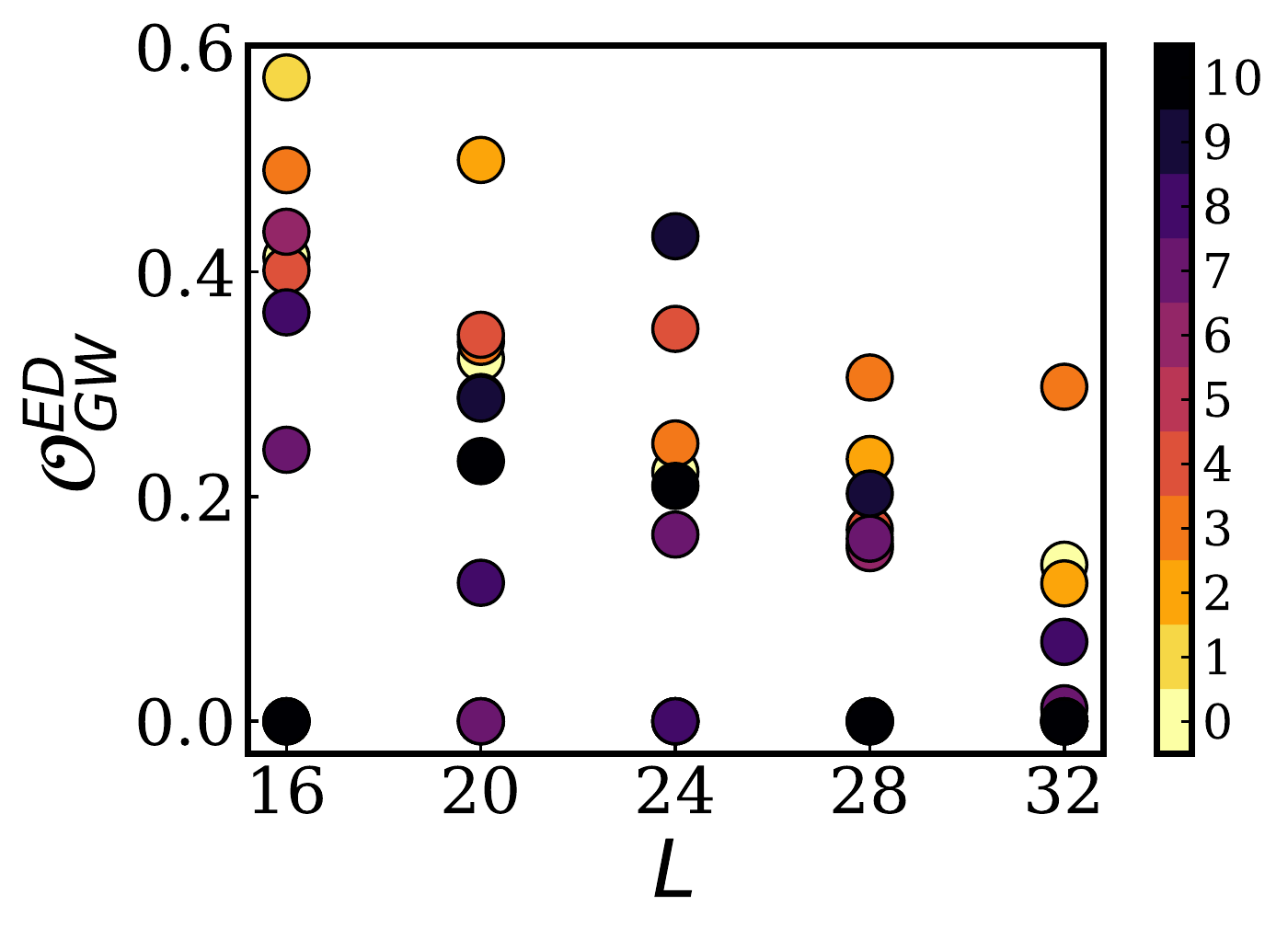}
    \put (45,73) {{\textbf{(a)}}}
    \end{overpic}
    \begin{overpic}[width=0.45\linewidth]{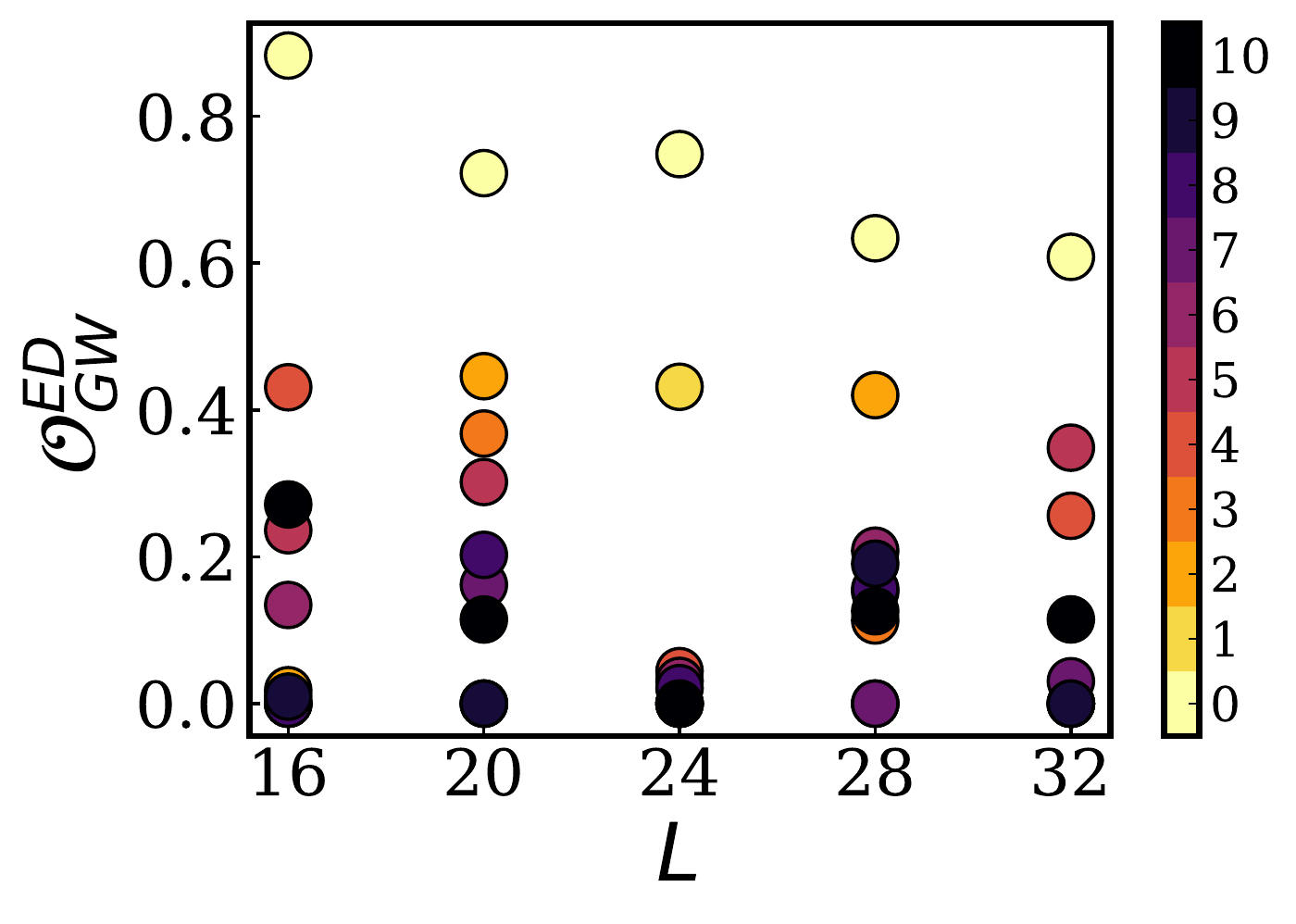}
    \put (45,73) {{\textbf{(b)}}}
    \end{overpic}
    \begin{overpic}[width=0.45\linewidth]{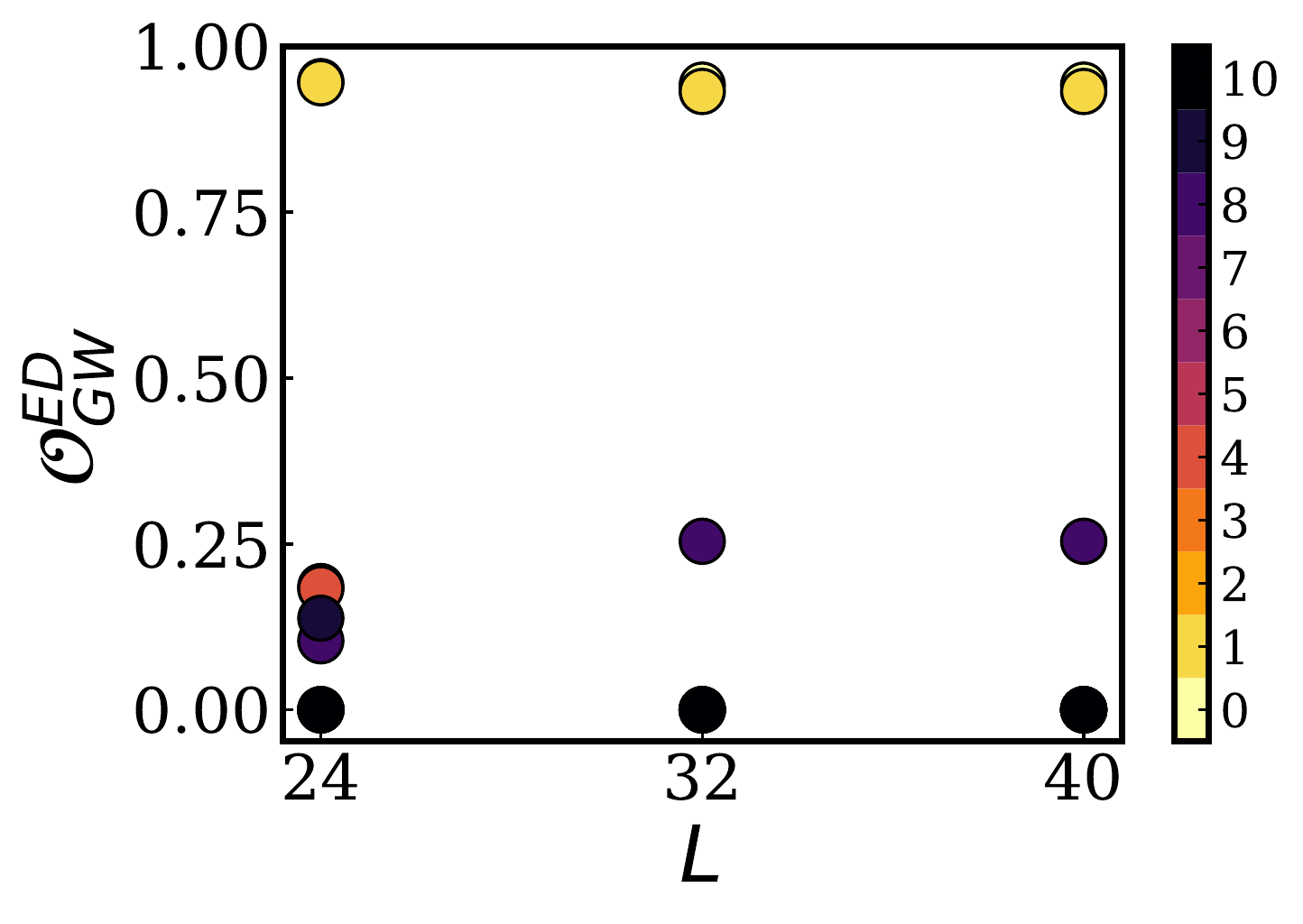}
    \put (45,73) {{\textbf{(c)}}}
    \end{overpic}
    \caption{Overlaps $\mathcal{O}^{ED}_{GW}$ with low-lying states in the $k=(0,0)$ momentum sector up to the 10th excited state at (a) $g=0.1$ and (b) $g=0.7$ for 1/2-density. (c) Overlaps for 1/4-density at $g=0.2$. The markers are colored according to the level of the state.}
    \label{fig:overlap2}
\end{figure}

In Fig. \ref{fig:overlap2}a,b, we present the overlaps $\mathcal{O}^{ED}_{GW}$ with the low-lying states in the $k=(0,0)$ momentum sector at $g=0.1$ and $g=0.7$. This provides additional insights into the behavior of the overlaps. It can be seen more clearly that at $g=0.7$, the overlap is the largest with the ground state. Furthermore, there is an additional low-lying state with a modest overlap, possibly representing the topological ground state in the thermodynamic limit. In contrast, such clear pattern is not observed at $g=0.1$. Instead, the overlaps are seen to be decreasing with system size.

In the case of 1/4-density, the approximate two-fold degeneracy can already be observed clearly from the ED spectra, even on small-size clusters. In addition, as seen in Fig. \ref{fig:overlap2}c, both of the nearly-degenerate states have huge overlaps with the optimal wavefunction ansatz. 

\section{Details on DMRG simulations} \label{app:dmrg}

We perform density-matrix renormalization group (DMRG)~\cite{white_prl_1992,white_prb_1993,schollwock_aop_2011, Orus_aop_2014} simulations using matrix-product state (MPS)~\cite{schollwock_aop_2011, Orus_aop_2014} anstaz for both finite and infinite cylinders, where we take periodic boundary conditions along the one axis with length $L_{\text{PBC}}=4$ (8 lattice sites). For finite-size cylinders, we make use of the ITensor library~\cite{itensor} with home-grown DMRG codes.
For infinite cylinders, we employ the TeNPy library~\cite{tenpy} that uses
 infinite MPS (iMPS)~\cite{Vidal_2007, Kjall_PRB_2013} ansatz for infinite
DMRG (iDMRG)~\cite{McCulloch_2008, Crosswhite_2008} simulations. The MPS bond-dimension for the simulations has been taken in the range $\chi \in [2000, 4000]$ resulting in cut-off errors of the order of $10^{-5}$ to $10^{-7}$ depending on the system parameters. For completeness, we analyze two different cylinder geometries, namely the geometries I and II as sketched in \cite{pollmann_prl_2015}.

\section{Transfer matrix spectrum} \label{app:transfer_matrix}

The nature of the excitation spectrum can be examined by looking at the spectrum of the transfer matrix  of the iMPS ansatz in the iDMRG simulations (see \cite{he2017,hu2019}). For that purpose, we inject a twist angle $\theta_{\text{PBC}}$ by putting a $\theta_{\text{PBC}}$-flux across the periodic direction cylinder, and follow the spectrum of the iMPS transfer matrix. It is to be noted that the iMPS correlation length $\xi_i$ of a low-energy excitation is related to  the transfer matrix eigenvalue   $\lambda_i$ as $\xi_i=-1/\ln|\lambda_i|$.

\begin{figure}
    \centering
\includegraphics[width=0.8\linewidth]{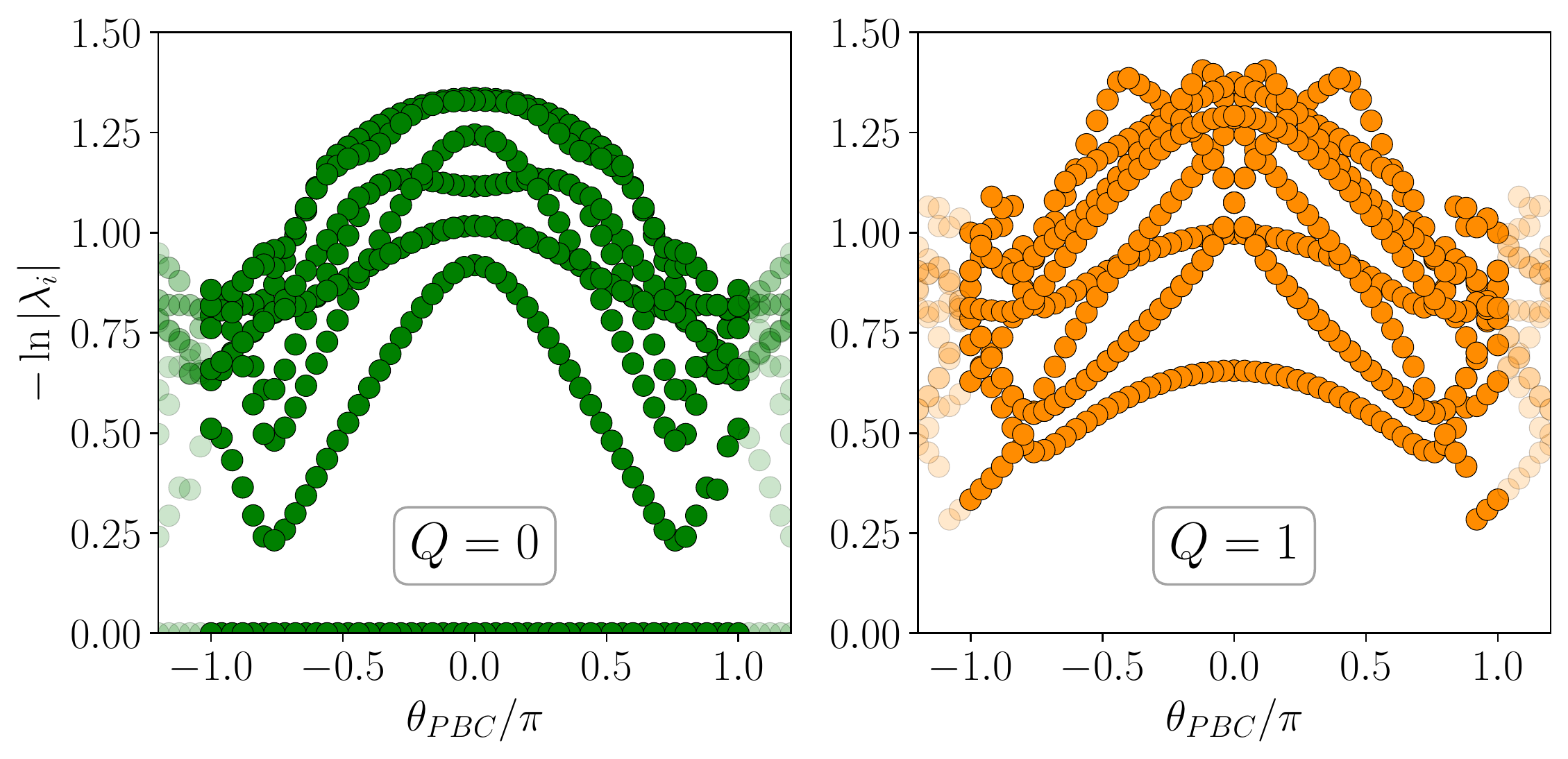}
    \caption{The spectrum of iMPS transfer matrix in the $Q=0$ and $Q=1$ sectors as functions of the twist angle $\theta_{\text{PBC}}$ for $g=0.74$ and density $n=1/2$.
    Here, we consider an infinite cylinder (geometry I) with periodic width $L_{\text{PBC}} = 4$ and iMPS bond dimension $\chi=4000$.
    }
\label{fig:transfer_matrix_spectrum}
\end{figure}

In Fig. \ref{fig:transfer_matrix_spectrum}, we show the low-lying iMPS transfer matrix spectrum in the  charge $Q=0$ and $Q=1$ sectors at $g=0.74$ as a function of the twist angle $\theta_{\text{PBC}}$ for an infinite cylinder (geometry I). 
Interestingly, for both $Q=0$ and $1$, the spectrum is not symmetric around $\theta_{\text{PBC}}=\pi$ which could be attributed to a remnant effect due to the absence of time-reversal symmetry. 
Nevertheless, the results do not show any signatures of a Dirac cone behavior, which would have appeared as linear dispersion around the minimum gap \cite{he2017,hu2019}. Instead, the transfer matrix spectrum is consistent with a Chern insulator, where the dispersion becomes quadratic around the minimum gap for $Q=0$. However, for $Q=1$, we see a sharp jump in the spectrum around $\theta_{\text{PBC}} \approx 0.9 \pi$ which could be a numerical artifact of finite iMPS bond dimension which we set here to $\chi = 4000$.

\bibliography{biblio}

\begin{thebibliography}{71}%
\makeatletter
\providecommand \@ifxundefined [1]{%
 \@ifx{#1\undefined}
}%
\providecommand \@ifnum [1]{%
 \ifnum #1\expandafter \@firstoftwo
 \else \expandafter \@secondoftwo
 \fi
}%
\providecommand \@ifx [1]{%
 \ifx #1\expandafter \@firstoftwo
 \else \expandafter \@secondoftwo
 \fi
}%
\providecommand \natexlab [1]{#1}%
\providecommand \enquote  [1]{``#1''}%
\providecommand \bibnamefont  [1]{#1}%
\providecommand \bibfnamefont [1]{#1}%
\providecommand \citenamefont [1]{#1}%
\providecommand \href@noop [0]{\@secondoftwo}%
\providecommand \href [0]{\begingroup \@sanitize@url \@href}%
\providecommand \@href[1]{\@@startlink{#1}\@@href}%
\providecommand \@@href[1]{\endgroup#1\@@endlink}%
\providecommand \@sanitize@url [0]{\catcode `\\12\catcode `\$12\catcode
  `\&12\catcode `\#12\catcode `\^12\catcode `\_12\catcode `\%12\relax}%
\providecommand \@@startlink[1]{}%
\providecommand \@@endlink[0]{}%
\providecommand \url  [0]{\begingroup\@sanitize@url \@url }%
\providecommand \@url [1]{\endgroup\@href {#1}{\urlprefix }}%
\providecommand \urlprefix  [0]{URL }%
\providecommand \Eprint [0]{\href }%
\providecommand \doibase [0]{https://doi.org/}%
\providecommand \selectlanguage [0]{\@gobble}%
\providecommand \bibinfo  [0]{\@secondoftwo}%
\providecommand \bibfield  [0]{\@secondoftwo}%
\providecommand \translation [1]{[#1]}%
\providecommand \BibitemOpen [0]{}%
\providecommand \bibitemStop [0]{}%
\providecommand \bibitemNoStop [0]{.\EOS\space}%
\providecommand \EOS [0]{\spacefactor3000\relax}%
\providecommand \BibitemShut  [1]{\csname bibitem#1\endcsname}%
\let\auto@bib@innerbib\@empty
\bibitem [{\citenamefont {Gross}\ and\ \citenamefont
  {Bloch}(2017)}]{gross2017quantum}%
  \BibitemOpen
  \bibfield  {author} {\bibinfo {author} {\bibfnamefont {C.}~\bibnamefont
  {Gross}}\ and\ \bibinfo {author} {\bibfnamefont {I.}~\bibnamefont {Bloch}},\
  }\bibfield  {title} {\bibinfo {title} {{Quantum simulations with ultracold
  atoms in optical lattices}},\ }\href
  {https://doi.org/10.1126/science.aal3837} {\bibfield  {journal} {\bibinfo
  {journal} {Science}\ }\textbf {\bibinfo {volume} {357}},\ \bibinfo {pages}
  {995} (\bibinfo {year} {2017})}\BibitemShut {NoStop}%
\bibitem [{\citenamefont {Browaeys}\ and\ \citenamefont
  {Lahaye}(2020)}]{browaeys_2020}%
  \BibitemOpen
  \bibfield  {author} {\bibinfo {author} {\bibfnamefont {A.}~\bibnamefont
  {Browaeys}}\ and\ \bibinfo {author} {\bibfnamefont {T.}~\bibnamefont
  {Lahaye}},\ }\bibfield  {title} {\bibinfo {title} {{Many-body physics with
  individually controlled Rydberg atoms}},\ }\href
  {https://doi.org/10.1038/s41567-019-0733-z} {\bibfield  {journal} {\bibinfo
  {journal} {Nature Physics}\ }\textbf {\bibinfo {volume} {16}},\ \bibinfo
  {pages} {132} (\bibinfo {year} {2020})}\BibitemShut {NoStop}%
\bibitem [{\citenamefont {de~L{\'{e}}s{\'{e}}leuc}\ \emph
  {et~al.}(2019)\citenamefont {de~L{\'{e}}s{\'{e}}leuc}, \citenamefont
  {Lienhard}, \citenamefont {Scholl}, \citenamefont {Barredo}, \citenamefont
  {Weber}, \citenamefont {Lang}, \citenamefont {B\"{u}chler}, \citenamefont
  {Lahaye},\ and\ \citenamefont {Browaeys}}]{SSHRyd2019}%
  \BibitemOpen
  \bibfield  {author} {\bibinfo {author} {\bibfnamefont {S.}~\bibnamefont
  {de~L{\'{e}}s{\'{e}}leuc}}, \bibinfo {author} {\bibfnamefont
  {V.}~\bibnamefont {Lienhard}}, \bibinfo {author} {\bibfnamefont
  {P.}~\bibnamefont {Scholl}}, \bibinfo {author} {\bibfnamefont
  {D.}~\bibnamefont {Barredo}}, \bibinfo {author} {\bibfnamefont
  {S.}~\bibnamefont {Weber}}, \bibinfo {author} {\bibfnamefont
  {N.}~\bibnamefont {Lang}}, \bibinfo {author} {\bibfnamefont {H.~P.}\
  \bibnamefont {B\"{u}chler}}, \bibinfo {author} {\bibfnamefont
  {T.}~\bibnamefont {Lahaye}},\ and\ \bibinfo {author} {\bibfnamefont
  {A.}~\bibnamefont {Browaeys}},\ }\bibfield  {title} {\bibinfo {title}
  {{Observation of a symmetry-protected topological phase of interacting bosons
  with Rydberg atoms}},\ }\href {https://doi.org/10.1126/science.aav9105}
  {\bibfield  {journal} {\bibinfo  {journal} {Science}\ }\textbf {\bibinfo
  {volume} {365}},\ \bibinfo {pages} {775} (\bibinfo {year}
  {2019})}\BibitemShut {NoStop}%
\bibitem [{\citenamefont {Scholl}\ \emph {et~al.}(2021)\citenamefont {Scholl},
  \citenamefont {Schuler}, \citenamefont {Williams}, \citenamefont
  {Eberharter}, \citenamefont {Barredo}, \citenamefont {Schymik}, \citenamefont
  {Lienhard}, \citenamefont {Henry}, \citenamefont {Lang}, \citenamefont
  {Lahaye}, \citenamefont {L\"{a}uchli},\ and\ \citenamefont
  {Browaeys}}]{scholl2021quantum}%
  \BibitemOpen
  \bibfield  {author} {\bibinfo {author} {\bibfnamefont {P.}~\bibnamefont
  {Scholl}}, \bibinfo {author} {\bibfnamefont {M.}~\bibnamefont {Schuler}},
  \bibinfo {author} {\bibfnamefont {H.~J.}\ \bibnamefont {Williams}}, \bibinfo
  {author} {\bibfnamefont {A.~A.}\ \bibnamefont {Eberharter}}, \bibinfo
  {author} {\bibfnamefont {D.}~\bibnamefont {Barredo}}, \bibinfo {author}
  {\bibfnamefont {K.-N.}\ \bibnamefont {Schymik}}, \bibinfo {author}
  {\bibfnamefont {V.}~\bibnamefont {Lienhard}}, \bibinfo {author}
  {\bibfnamefont {L.-P.}\ \bibnamefont {Henry}}, \bibinfo {author}
  {\bibfnamefont {T.~C.}\ \bibnamefont {Lang}}, \bibinfo {author}
  {\bibfnamefont {T.}~\bibnamefont {Lahaye}}, \bibinfo {author} {\bibfnamefont
  {A.~M.}\ \bibnamefont {L\"{a}uchli}},\ and\ \bibinfo {author} {\bibfnamefont
  {A.}~\bibnamefont {Browaeys}},\ }\bibfield  {title} {\bibinfo {title}
  {{Quantum simulation of 2D antiferromagnets with hundreds of Rydberg
  atoms}},\ }\href {https://doi.org/10.1038/s41586-021-03585-1} {\bibfield
  {journal} {\bibinfo  {journal} {Nature}\ }\textbf {\bibinfo {volume} {595}},\
  \bibinfo {pages} {233} (\bibinfo {year} {2021})}\BibitemShut {NoStop}%
\bibitem [{\citenamefont {Ebadi}\ \emph {et~al.}(2021)\citenamefont {Ebadi},
  \citenamefont {Wang}, \citenamefont {Levine}, \citenamefont {Keesling},
  \citenamefont {Semeghini}, \citenamefont {Omran}, \citenamefont {Bluvstein},
  \citenamefont {Samajdar}, \citenamefont {Pichler}, \citenamefont {Ho},
  \citenamefont {Choi}, \citenamefont {Sachdev}, \citenamefont {Greiner},
  \citenamefont {Vuleti{\'{c}}},\ and\ \citenamefont
  {Lukin}}]{ebadi2021quantum}%
  \BibitemOpen
  \bibfield  {author} {\bibinfo {author} {\bibfnamefont {S.}~\bibnamefont
  {Ebadi}}, \bibinfo {author} {\bibfnamefont {T.~T.}\ \bibnamefont {Wang}},
  \bibinfo {author} {\bibfnamefont {H.}~\bibnamefont {Levine}}, \bibinfo
  {author} {\bibfnamefont {A.}~\bibnamefont {Keesling}}, \bibinfo {author}
  {\bibfnamefont {G.}~\bibnamefont {Semeghini}}, \bibinfo {author}
  {\bibfnamefont {A.}~\bibnamefont {Omran}}, \bibinfo {author} {\bibfnamefont
  {D.}~\bibnamefont {Bluvstein}}, \bibinfo {author} {\bibfnamefont
  {R.}~\bibnamefont {Samajdar}}, \bibinfo {author} {\bibfnamefont
  {H.}~\bibnamefont {Pichler}}, \bibinfo {author} {\bibfnamefont {W.~W.}\
  \bibnamefont {Ho}}, \bibinfo {author} {\bibfnamefont {S.}~\bibnamefont
  {Choi}}, \bibinfo {author} {\bibfnamefont {S.}~\bibnamefont {Sachdev}},
  \bibinfo {author} {\bibfnamefont {M.}~\bibnamefont {Greiner}}, \bibinfo
  {author} {\bibfnamefont {V.}~\bibnamefont {Vuleti{\'{c}}}},\ and\ \bibinfo
  {author} {\bibfnamefont {M.~D.}\ \bibnamefont {Lukin}},\ }\bibfield  {title}
  {\bibinfo {title} {{Quantum phases of matter on a 256-atom programmable
  quantum simulator}},\ }\href {https://doi.org/10.1038/s41586-021-03582-4}
  {\bibfield  {journal} {\bibinfo  {journal} {Nature}\ }\textbf {\bibinfo
  {volume} {595}},\ \bibinfo {pages} {227} (\bibinfo {year}
  {2021})}\BibitemShut {NoStop}%
\bibitem [{\citenamefont {Semeghini}\ \emph {et~al.}(2021)\citenamefont
  {Semeghini}, \citenamefont {Levine}, \citenamefont {Keesling}, \citenamefont
  {Ebadi}, \citenamefont {Wang}, \citenamefont {Bluvstein}, \citenamefont
  {Verresen}, \citenamefont {Pichler}, \citenamefont {Kalinowski},
  \citenamefont {Samajdar}, \citenamefont {Omran}, \citenamefont {Sachdev},
  \citenamefont {Vishwanath}, \citenamefont {Greiner}, \citenamefont
  {Vuleti{\'{c}}},\ and\ \citenamefont {Lukin}}]{Semeghini2021}%
  \BibitemOpen
  \bibfield  {author} {\bibinfo {author} {\bibfnamefont {G.}~\bibnamefont
  {Semeghini}}, \bibinfo {author} {\bibfnamefont {H.}~\bibnamefont {Levine}},
  \bibinfo {author} {\bibfnamefont {A.}~\bibnamefont {Keesling}}, \bibinfo
  {author} {\bibfnamefont {S.}~\bibnamefont {Ebadi}}, \bibinfo {author}
  {\bibfnamefont {T.~T.}\ \bibnamefont {Wang}}, \bibinfo {author}
  {\bibfnamefont {D.}~\bibnamefont {Bluvstein}}, \bibinfo {author}
  {\bibfnamefont {R.}~\bibnamefont {Verresen}}, \bibinfo {author}
  {\bibfnamefont {H.}~\bibnamefont {Pichler}}, \bibinfo {author} {\bibfnamefont
  {M.}~\bibnamefont {Kalinowski}}, \bibinfo {author} {\bibfnamefont
  {R.}~\bibnamefont {Samajdar}}, \bibinfo {author} {\bibfnamefont
  {A.}~\bibnamefont {Omran}}, \bibinfo {author} {\bibfnamefont
  {S.}~\bibnamefont {Sachdev}}, \bibinfo {author} {\bibfnamefont
  {A.}~\bibnamefont {Vishwanath}}, \bibinfo {author} {\bibfnamefont
  {M.}~\bibnamefont {Greiner}}, \bibinfo {author} {\bibfnamefont
  {V.}~\bibnamefont {Vuleti{\'{c}}}},\ and\ \bibinfo {author} {\bibfnamefont
  {M.~D.}\ \bibnamefont {Lukin}},\ }\bibfield  {title} {\bibinfo {title}
  {{Probing topological spin liquids on a programmable quantum simulator}},\
  }\href {https://doi.org/10.1126/science.abi8794} {\bibfield  {journal}
  {\bibinfo  {journal} {Science}\ }\textbf {\bibinfo {volume} {374}},\ \bibinfo
  {pages} {1242} (\bibinfo {year} {2021})}\BibitemShut {NoStop}%
\bibitem [{\citenamefont {Savary}\ and\ \citenamefont
  {Balents}(2016)}]{Savary_2016}%
  \BibitemOpen
  \bibfield  {author} {\bibinfo {author} {\bibfnamefont {L.}~\bibnamefont
  {Savary}}\ and\ \bibinfo {author} {\bibfnamefont {L.}~\bibnamefont
  {Balents}},\ }\bibfield  {title} {\bibinfo {title} {{Quantum spin liquids: a
  review}},\ }\href {https://doi.org/10.1088/0034-4885/80/1/016502} {\bibfield
  {journal} {\bibinfo  {journal} {Reports on Progress in Physics}\ }\textbf
  {\bibinfo {volume} {80}},\ \bibinfo {pages} {016502} (\bibinfo {year}
  {2016})}\BibitemShut {NoStop}%
\bibitem [{\citenamefont {Wen}(2017)}]{Wen_2017}%
  \BibitemOpen
  \bibfield  {author} {\bibinfo {author} {\bibfnamefont {X.-G.}\ \bibnamefont
  {Wen}},\ }\bibfield  {title} {\bibinfo {title} {{Colloquium: Zoo of
  quantum-topological phases of matter}},\ }\href
  {https://doi.org/10.1103/RevModPhys.89.041004} {\bibfield  {journal}
  {\bibinfo  {journal} {Rev. Mod. Phys.}\ }\textbf {\bibinfo {volume} {89}},\
  \bibinfo {pages} {041004} (\bibinfo {year} {2017})}\BibitemShut {NoStop}%
\bibitem [{\citenamefont {Lacroix}\ \emph {et~al.}(2011)\citenamefont
  {Lacroix}, \citenamefont {Mendels},\ and\ \citenamefont
  {Mila}}]{lacroix2011}%
  \BibitemOpen
  \bibinfo {editor} {\bibfnamefont {C.}~\bibnamefont {Lacroix}}, \bibinfo
  {editor} {\bibfnamefont {P.}~\bibnamefont {Mendels}},\ and\ \bibinfo {editor}
  {\bibfnamefont {F.}~\bibnamefont {Mila}},\ eds.,\ \href
  {https://doi.org/10.1007/978-3-642-10589-0} {\emph {\bibinfo {title}
  {{Introduction to Frustrated Magnetism}}}}\ (\bibinfo  {publisher} {Springer
  Berlin Heidelberg},\ \bibinfo {year} {2011})\BibitemShut {NoStop}%
\bibitem [{\citenamefont {Moessner}\ and\ \citenamefont
  {Moore}(2021)}]{moessner_moore_2021}%
  \BibitemOpen
  \bibfield  {author} {\bibinfo {author} {\bibfnamefont {R.}~\bibnamefont
  {Moessner}}\ and\ \bibinfo {author} {\bibfnamefont {J.~E.}\ \bibnamefont
  {Moore}},\ }\href {https://doi.org/10.1017/9781316226308} {\emph {\bibinfo
  {title} {{Topological Phases of Matter}}}}\ (\bibinfo  {publisher} {Cambridge
  University Press},\ \bibinfo {year} {2021})\BibitemShut {NoStop}%
\bibitem [{\citenamefont {Lukin}\ \emph {et~al.}(2001)\citenamefont {Lukin},
  \citenamefont {Fleischhauer}, \citenamefont {Cote}, \citenamefont {Duan},
  \citenamefont {Jaksch}, \citenamefont {Cirac},\ and\ \citenamefont
  {Zoller}}]{lukin2001dipole}%
  \BibitemOpen
  \bibfield  {author} {\bibinfo {author} {\bibfnamefont {M.~D.}\ \bibnamefont
  {Lukin}}, \bibinfo {author} {\bibfnamefont {M.}~\bibnamefont {Fleischhauer}},
  \bibinfo {author} {\bibfnamefont {R.}~\bibnamefont {Cote}}, \bibinfo {author}
  {\bibfnamefont {L.~M.}\ \bibnamefont {Duan}}, \bibinfo {author}
  {\bibfnamefont {D.}~\bibnamefont {Jaksch}}, \bibinfo {author} {\bibfnamefont
  {J.~I.}\ \bibnamefont {Cirac}},\ and\ \bibinfo {author} {\bibfnamefont
  {P.}~\bibnamefont {Zoller}},\ }\bibfield  {title} {\bibinfo {title} {{Dipole
  Blockade and Quantum Information Processing in Mesoscopic Atomic
  Ensembles}},\ }\href {https://doi.org/10.1103/PhysRevLett.87.037901}
  {\bibfield  {journal} {\bibinfo  {journal} {Phys. Rev. Lett.}\ }\textbf
  {\bibinfo {volume} {87}},\ \bibinfo {pages} {037901} (\bibinfo {year}
  {2001})}\BibitemShut {NoStop}%
\bibitem [{\citenamefont {Schauss}(2018)}]{schauss2018quantum}%
  \BibitemOpen
  \bibfield  {author} {\bibinfo {author} {\bibfnamefont {P.}~\bibnamefont
  {Schauss}},\ }\bibfield  {title} {\bibinfo {title} {{Quantum simulation of
  transverse Ising models with Rydberg atoms}},\ }\href
  {https://doi.org/10.1088/2058-9565/aa9c59} {\bibfield  {journal} {\bibinfo
  {journal} {Quantum Science and Technology}\ }\textbf {\bibinfo {volume}
  {3}},\ \bibinfo {pages} {023001} (\bibinfo {year} {2018})}\BibitemShut
  {NoStop}%
\bibitem [{\citenamefont {Samajdar}\ \emph {et~al.}(2021)\citenamefont
  {Samajdar}, \citenamefont {Ho}, \citenamefont {Pichler}, \citenamefont
  {Lukin},\ and\ \citenamefont {Sachdev}}]{Samajdar2021}%
  \BibitemOpen
  \bibfield  {author} {\bibinfo {author} {\bibfnamefont {R.}~\bibnamefont
  {Samajdar}}, \bibinfo {author} {\bibfnamefont {W.~W.}\ \bibnamefont {Ho}},
  \bibinfo {author} {\bibfnamefont {H.}~\bibnamefont {Pichler}}, \bibinfo
  {author} {\bibfnamefont {M.~D.}\ \bibnamefont {Lukin}},\ and\ \bibinfo
  {author} {\bibfnamefont {S.}~\bibnamefont {Sachdev}},\ }\bibfield  {title}
  {\bibinfo {title} {{Quantum phases of Rydberg atoms on a kagome lattice}},\
  }\href {https://doi.org/10.1073/pnas.2015785118} {\bibfield  {journal}
  {\bibinfo  {journal} {Proceedings of the National Academy of Sciences}\
  }\textbf {\bibinfo {volume} {118}},\ \bibinfo {pages} {e2015785118} (\bibinfo
  {year} {2021})}\BibitemShut {NoStop}%
\bibitem [{\citenamefont {Verresen}\ \emph {et~al.}(2021)\citenamefont
  {Verresen}, \citenamefont {Lukin},\ and\ \citenamefont
  {Vishwanath}}]{Verresen2021}%
  \BibitemOpen
  \bibfield  {author} {\bibinfo {author} {\bibfnamefont {R.}~\bibnamefont
  {Verresen}}, \bibinfo {author} {\bibfnamefont {M.~D.}\ \bibnamefont
  {Lukin}},\ and\ \bibinfo {author} {\bibfnamefont {A.}~\bibnamefont
  {Vishwanath}},\ }\bibfield  {title} {\bibinfo {title} {{Prediction of Toric
  Code Topological Order from Rydberg Blockade}},\ }\href
  {https://doi.org/10.1103/PhysRevX.11.031005} {\bibfield  {journal} {\bibinfo
  {journal} {Phys. Rev. X}\ }\textbf {\bibinfo {volume} {11}},\ \bibinfo
  {pages} {031005} (\bibinfo {year} {2021})}\BibitemShut {NoStop}%
\bibitem [{\citenamefont {Giudici}\ \emph {et~al.}(2022)\citenamefont
  {Giudici}, \citenamefont {Lukin},\ and\ \citenamefont
  {Pichler}}]{giudici2022}%
  \BibitemOpen
  \bibfield  {author} {\bibinfo {author} {\bibfnamefont {G.}~\bibnamefont
  {Giudici}}, \bibinfo {author} {\bibfnamefont {M.~D.}\ \bibnamefont {Lukin}},\
  and\ \bibinfo {author} {\bibfnamefont {H.}~\bibnamefont {Pichler}},\
  }\bibfield  {title} {\bibinfo {title} {{Dynamical Preparation of Quantum Spin
  Liquids in Rydberg Atom Arrays}},\ }\href
  {https://doi.org/10.1103/PhysRevLett.129.090401} {\bibfield  {journal}
  {\bibinfo  {journal} {Phys. Rev. Lett.}\ }\textbf {\bibinfo {volume} {129}},\
  \bibinfo {pages} {090401} (\bibinfo {year} {2022})}\BibitemShut {NoStop}%
\bibitem [{\citenamefont {Giudice}\ \emph {et~al.}(2022)\citenamefont
  {Giudice}, \citenamefont {Surace}, \citenamefont {Pichler},\ and\
  \citenamefont {Giudici}}]{Giudice_2022}%
  \BibitemOpen
  \bibfield  {author} {\bibinfo {author} {\bibfnamefont {G.}~\bibnamefont
  {Giudice}}, \bibinfo {author} {\bibfnamefont {F.~M.}\ \bibnamefont {Surace}},
  \bibinfo {author} {\bibfnamefont {H.}~\bibnamefont {Pichler}},\ and\ \bibinfo
  {author} {\bibfnamefont {G.}~\bibnamefont {Giudici}},\ }\bibfield  {title}
  {\bibinfo {title} {{Trimer states with $\mathbb{Z}_{3}$ topological order in
  Rydberg atom arrays}},\ }\href {https://doi.org/10.1103/PhysRevB.106.195155}
  {\bibfield  {journal} {\bibinfo  {journal} {Phys. Rev. B}\ }\textbf {\bibinfo
  {volume} {106}},\ \bibinfo {pages} {195155} (\bibinfo {year}
  {2022})}\BibitemShut {NoStop}%
\bibitem [{\citenamefont {Moessner}\ and\ \citenamefont
  {Sondhi}(2001)}]{moessner2001ising}%
  \BibitemOpen
  \bibfield  {author} {\bibinfo {author} {\bibfnamefont {R.}~\bibnamefont
  {Moessner}}\ and\ \bibinfo {author} {\bibfnamefont {S.~L.}\ \bibnamefont
  {Sondhi}},\ }\bibfield  {title} {\bibinfo {title} {{Ising models of quantum
  frustration}},\ }\href {https://doi.org/10.1103/PhysRevB.63.224401}
  {\bibfield  {journal} {\bibinfo  {journal} {Phys. Rev. B}\ }\textbf {\bibinfo
  {volume} {63}},\ \bibinfo {pages} {224401} (\bibinfo {year}
  {2001})}\BibitemShut {NoStop}%
\bibitem [{\citenamefont {Glaetzle}\ \emph {et~al.}(2014)\citenamefont
  {Glaetzle}, \citenamefont {Dalmonte}, \citenamefont {Nath}, \citenamefont
  {Rousochatzakis}, \citenamefont {Moessner},\ and\ \citenamefont
  {Zoller}}]{glaetzle2014quantum}%
  \BibitemOpen
  \bibfield  {author} {\bibinfo {author} {\bibfnamefont {A.~W.}\ \bibnamefont
  {Glaetzle}}, \bibinfo {author} {\bibfnamefont {M.}~\bibnamefont {Dalmonte}},
  \bibinfo {author} {\bibfnamefont {R.}~\bibnamefont {Nath}}, \bibinfo {author}
  {\bibfnamefont {I.}~\bibnamefont {Rousochatzakis}}, \bibinfo {author}
  {\bibfnamefont {R.}~\bibnamefont {Moessner}},\ and\ \bibinfo {author}
  {\bibfnamefont {P.}~\bibnamefont {Zoller}},\ }\bibfield  {title} {\bibinfo
  {title} {{Quantum Spin-Ice and Dimer Models with Rydberg Atoms}},\ }\href
  {https://doi.org/10.1103/PhysRevX.4.041037} {\bibfield  {journal} {\bibinfo
  {journal} {Phys. Rev. X}\ }\textbf {\bibinfo {volume} {4}},\ \bibinfo {pages}
  {041037} (\bibinfo {year} {2014})}\BibitemShut {NoStop}%
\bibitem [{\citenamefont {Tarabunga}\ \emph {et~al.}(2022)\citenamefont
  {Tarabunga}, \citenamefont {Surace}, \citenamefont {Andreoni}, \citenamefont
  {Angelone},\ and\ \citenamefont {Dalmonte}}]{Tarabunga_2022}%
  \BibitemOpen
  \bibfield  {author} {\bibinfo {author} {\bibfnamefont {P.~S.}\ \bibnamefont
  {Tarabunga}}, \bibinfo {author} {\bibfnamefont {F.~M.}\ \bibnamefont
  {Surace}}, \bibinfo {author} {\bibfnamefont {R.}~\bibnamefont {Andreoni}},
  \bibinfo {author} {\bibfnamefont {A.}~\bibnamefont {Angelone}},\ and\
  \bibinfo {author} {\bibfnamefont {M.}~\bibnamefont {Dalmonte}},\ }\bibfield
  {title} {\bibinfo {title} {{Gauge-Theoretic Origin of Rydberg Quantum Spin
  Liquids}},\ }\href {https://doi.org/10.1103/PhysRevLett.129.195301}
  {\bibfield  {journal} {\bibinfo  {journal} {Phys. Rev. Lett.}\ }\textbf
  {\bibinfo {volume} {129}},\ \bibinfo {pages} {195301} (\bibinfo {year}
  {2022})}\BibitemShut {NoStop}%
\bibitem [{\citenamefont {Samajdar}\ \emph {et~al.}(2023)\citenamefont
  {Samajdar}, \citenamefont {Joshi}, \citenamefont {Teng},\ and\ \citenamefont
  {Sachdev}}]{Samajdar2022}%
  \BibitemOpen
  \bibfield  {author} {\bibinfo {author} {\bibfnamefont {R.}~\bibnamefont
  {Samajdar}}, \bibinfo {author} {\bibfnamefont {D.~G.}\ \bibnamefont {Joshi}},
  \bibinfo {author} {\bibfnamefont {Y.}~\bibnamefont {Teng}},\ and\ \bibinfo
  {author} {\bibfnamefont {S.}~\bibnamefont {Sachdev}},\ }\bibfield  {title}
  {\bibinfo {title} {{Emergent $\mathbb{Z}_{2}$ Gauge Theories and Topological
  Excitations in Rydberg Atom Arrays}},\ }\href
  {https://doi.org/10.1103/PhysRevLett.130.043601} {\bibfield  {journal}
  {\bibinfo  {journal} {Phys. Rev. Lett.}\ }\textbf {\bibinfo {volume} {130}},\
  \bibinfo {pages} {043601} (\bibinfo {year} {2023})}\BibitemShut {NoStop}%
\bibitem [{\citenamefont {Cheng}\ \emph {et~al.}(2023)\citenamefont {Cheng},
  \citenamefont {Li},\ and\ \citenamefont {Zhai}}]{Cheng2023}%
  \BibitemOpen
  \bibfield  {author} {\bibinfo {author} {\bibfnamefont {Y.}~\bibnamefont
  {Cheng}}, \bibinfo {author} {\bibfnamefont {C.}~\bibnamefont {Li}},\ and\
  \bibinfo {author} {\bibfnamefont {H.}~\bibnamefont {Zhai}},\ }\bibfield
  {title} {\bibinfo {title} {Variational approach to quantum spin liquid in a
  rydberg atom simulator},\ }\href {https://doi.org/10.1088/1367-2630/acc125}
  {\bibfield  {journal} {\bibinfo  {journal} {New Journal of Physics}\ }\textbf
  {\bibinfo {volume} {25}},\ \bibinfo {pages} {033010} (\bibinfo {year}
  {2023})}\BibitemShut {NoStop}%
\bibitem [{\citenamefont {Peter}\ \emph {et~al.}(2015)\citenamefont {Peter},
  \citenamefont {Yao}, \citenamefont {Lang}, \citenamefont {Huber},
  \citenamefont {Lukin},\ and\ \citenamefont {B\"uchler}}]{peter_2015}%
  \BibitemOpen
  \bibfield  {author} {\bibinfo {author} {\bibfnamefont {D.}~\bibnamefont
  {Peter}}, \bibinfo {author} {\bibfnamefont {N.~Y.}\ \bibnamefont {Yao}},
  \bibinfo {author} {\bibfnamefont {N.}~\bibnamefont {Lang}}, \bibinfo {author}
  {\bibfnamefont {S.~D.}\ \bibnamefont {Huber}}, \bibinfo {author}
  {\bibfnamefont {M.~D.}\ \bibnamefont {Lukin}},\ and\ \bibinfo {author}
  {\bibfnamefont {H.~P.}\ \bibnamefont {B\"uchler}},\ }\bibfield  {title}
  {\bibinfo {title} {{Topological bands with a Chern number $C=2$ by dipolar
  exchange interactions}},\ }\href {https://doi.org/10.1103/PhysRevA.91.053617}
  {\bibfield  {journal} {\bibinfo  {journal} {Phys. Rev. A}\ }\textbf {\bibinfo
  {volume} {91}},\ \bibinfo {pages} {053617} (\bibinfo {year}
  {2015})}\BibitemShut {NoStop}%
\bibitem [{\citenamefont {Weber}\ \emph {et~al.}(2018)\citenamefont {Weber},
  \citenamefont {de~L{\'{e}}s{\'{e}}leuc}, \citenamefont {Lienhard},
  \citenamefont {Barredo}, \citenamefont {Lahaye}, \citenamefont {Browaeys},\
  and\ \citenamefont {B\"{u}chler}}]{Weber_2018}%
  \BibitemOpen
  \bibfield  {author} {\bibinfo {author} {\bibfnamefont {S.}~\bibnamefont
  {Weber}}, \bibinfo {author} {\bibfnamefont {S.}~\bibnamefont
  {de~L{\'{e}}s{\'{e}}leuc}}, \bibinfo {author} {\bibfnamefont
  {V.}~\bibnamefont {Lienhard}}, \bibinfo {author} {\bibfnamefont
  {D.}~\bibnamefont {Barredo}}, \bibinfo {author} {\bibfnamefont
  {T.}~\bibnamefont {Lahaye}}, \bibinfo {author} {\bibfnamefont
  {A.}~\bibnamefont {Browaeys}},\ and\ \bibinfo {author} {\bibfnamefont
  {H.~P.}\ \bibnamefont {B\"{u}chler}},\ }\bibfield  {title} {\bibinfo {title}
  {{Topologically protected edge states in small Rydberg systems}},\ }\href
  {https://doi.org/10.1088/2058-9565/aaca47} {\bibfield  {journal} {\bibinfo
  {journal} {Quantum Science and Technology}\ }\textbf {\bibinfo {volume}
  {3}},\ \bibinfo {pages} {044001} (\bibinfo {year} {2018})}\BibitemShut
  {NoStop}%
\bibitem [{\citenamefont {Weber}\ \emph {et~al.}(2022)\citenamefont {Weber},
  \citenamefont {Bai}, \citenamefont {Makki}, \citenamefont {M\"ogerle},
  \citenamefont {Lahaye}, \citenamefont {Browaeys}, \citenamefont {Daghofer},
  \citenamefont {Lang},\ and\ \citenamefont {B\"uchler}}]{Weber_2022}%
  \BibitemOpen
  \bibfield  {author} {\bibinfo {author} {\bibfnamefont {S.}~\bibnamefont
  {Weber}}, \bibinfo {author} {\bibfnamefont {R.}~\bibnamefont {Bai}}, \bibinfo
  {author} {\bibfnamefont {N.}~\bibnamefont {Makki}}, \bibinfo {author}
  {\bibfnamefont {J.}~\bibnamefont {M\"ogerle}}, \bibinfo {author}
  {\bibfnamefont {T.}~\bibnamefont {Lahaye}}, \bibinfo {author} {\bibfnamefont
  {A.}~\bibnamefont {Browaeys}}, \bibinfo {author} {\bibfnamefont
  {M.}~\bibnamefont {Daghofer}}, \bibinfo {author} {\bibfnamefont
  {N.}~\bibnamefont {Lang}},\ and\ \bibinfo {author} {\bibfnamefont {H.~P.}\
  \bibnamefont {B\"uchler}},\ }\bibfield  {title} {\bibinfo {title}
  {{Experimentally Accessible Scheme for a Fractional Chern Insulator in
  Rydberg Atoms}},\ }\href {https://doi.org/10.1103/PRXQuantum.3.030302}
  {\bibfield  {journal} {\bibinfo  {journal} {PRX Quantum}\ }\textbf {\bibinfo
  {volume} {3}},\ \bibinfo {pages} {030302} (\bibinfo {year}
  {2022})}\BibitemShut {NoStop}%
\bibitem [{\citenamefont {Ohler}\ \emph {et~al.}(2023)\citenamefont {Ohler},
  \citenamefont {Kiefer-Emmanouilidis},\ and\ \citenamefont
  {Fleischhauer}}]{Ohler_2022}%
  \BibitemOpen
  \bibfield  {author} {\bibinfo {author} {\bibfnamefont {S.}~\bibnamefont
  {Ohler}}, \bibinfo {author} {\bibfnamefont {M.}~\bibnamefont
  {Kiefer-Emmanouilidis}},\ and\ \bibinfo {author} {\bibfnamefont
  {M.}~\bibnamefont {Fleischhauer}},\ }\bibfield  {title} {\bibinfo {title}
  {Quantum spin liquids of rydberg excitations in a honeycomb lattice induced
  by density-dependent peierls phases},\ }\href
  {https://doi.org/10.1103/PhysRevResearch.5.013157} {\bibfield  {journal}
  {\bibinfo  {journal} {Phys. Rev. Res.}\ }\textbf {\bibinfo {volume} {5}},\
  \bibinfo {pages} {013157} (\bibinfo {year} {2023})}\BibitemShut {NoStop}%
\bibitem [{\citenamefont {Lienhard}\ \emph {et~al.}(2020)\citenamefont
  {Lienhard}, \citenamefont {Scholl}, \citenamefont {Weber}, \citenamefont
  {Barredo}, \citenamefont {de~L\'es\'eleuc}, \citenamefont {Bai},
  \citenamefont {Lang}, \citenamefont {Fleischhauer}, \citenamefont
  {B\"uchler}, \citenamefont {Lahaye},\ and\ \citenamefont
  {Browaeys}}]{Lienhard_2020}%
  \BibitemOpen
  \bibfield  {author} {\bibinfo {author} {\bibfnamefont {V.}~\bibnamefont
  {Lienhard}}, \bibinfo {author} {\bibfnamefont {P.}~\bibnamefont {Scholl}},
  \bibinfo {author} {\bibfnamefont {S.}~\bibnamefont {Weber}}, \bibinfo
  {author} {\bibfnamefont {D.}~\bibnamefont {Barredo}}, \bibinfo {author}
  {\bibfnamefont {S.}~\bibnamefont {de~L\'es\'eleuc}}, \bibinfo {author}
  {\bibfnamefont {R.}~\bibnamefont {Bai}}, \bibinfo {author} {\bibfnamefont
  {N.}~\bibnamefont {Lang}}, \bibinfo {author} {\bibfnamefont {M.}~\bibnamefont
  {Fleischhauer}}, \bibinfo {author} {\bibfnamefont {H.~P.}\ \bibnamefont
  {B\"uchler}}, \bibinfo {author} {\bibfnamefont {T.}~\bibnamefont {Lahaye}},\
  and\ \bibinfo {author} {\bibfnamefont {A.}~\bibnamefont {Browaeys}},\
  }\bibfield  {title} {\bibinfo {title} {{Realization of a Density-Dependent
  Peierls Phase in a Synthetic, Spin-Orbit Coupled Rydberg System}},\ }\href
  {https://doi.org/10.1103/PhysRevX.10.021031} {\bibfield  {journal} {\bibinfo
  {journal} {Phys. Rev. X}\ }\textbf {\bibinfo {volume} {10}},\ \bibinfo
  {pages} {021031} (\bibinfo {year} {2020})}\BibitemShut {NoStop}%
\bibitem [{\citenamefont {Wen}(2002{\natexlab{a}})}]{Wen_2002}%
  \BibitemOpen
  \bibfield  {author} {\bibinfo {author} {\bibfnamefont {X.-G.}\ \bibnamefont
  {Wen}},\ }\bibfield  {title} {\bibinfo {title} {{Quantum orders and symmetric
  spin liquids}},\ }\href {https://doi.org/10.1103/PhysRevB.65.165113}
  {\bibfield  {journal} {\bibinfo  {journal} {Phys. Rev. B}\ }\textbf {\bibinfo
  {volume} {65}},\ \bibinfo {pages} {165113} (\bibinfo {year}
  {2002}{\natexlab{a}})}\BibitemShut {NoStop}%
\bibitem [{\citenamefont {Wen}(2002{\natexlab{b}})}]{Wen_2002_2}%
  \BibitemOpen
  \bibfield  {author} {\bibinfo {author} {\bibfnamefont {X.-G.}\ \bibnamefont
  {Wen}},\ }\bibfield  {title} {\bibinfo {title} {{Quantum order: a quantum
  entanglement of many particles}},\ }\href
  {https://doi.org/10.1016/s0375-9601(02)00808-3} {\bibfield  {journal}
  {\bibinfo  {journal} {Physics Letters A}\ }\textbf {\bibinfo {volume}
  {300}},\ \bibinfo {pages} {175} (\bibinfo {year}
  {2002}{\natexlab{b}})}\BibitemShut {NoStop}%
\bibitem [{\citenamefont {Bieri}\ \emph {et~al.}(2016)\citenamefont {Bieri},
  \citenamefont {Lhuillier},\ and\ \citenamefont {Messio}}]{Bieri_2016}%
  \BibitemOpen
  \bibfield  {author} {\bibinfo {author} {\bibfnamefont {S.}~\bibnamefont
  {Bieri}}, \bibinfo {author} {\bibfnamefont {C.}~\bibnamefont {Lhuillier}},\
  and\ \bibinfo {author} {\bibfnamefont {L.}~\bibnamefont {Messio}},\
  }\bibfield  {title} {\bibinfo {title} {{Projective symmetry group
  classification of chiral spin liquids}},\ }\href
  {https://doi.org/10.1103/PhysRevB.93.094437} {\bibfield  {journal} {\bibinfo
  {journal} {Phys. Rev. B}\ }\textbf {\bibinfo {volume} {93}},\ \bibinfo
  {pages} {094437} (\bibinfo {year} {2016})}\BibitemShut {NoStop}%
\bibitem [{\citenamefont {Baskaran}\ and\ \citenamefont
  {Anderson}(1988)}]{Baskaran_1988}%
  \BibitemOpen
  \bibfield  {author} {\bibinfo {author} {\bibfnamefont {G.}~\bibnamefont
  {Baskaran}}\ and\ \bibinfo {author} {\bibfnamefont {P.~W.}\ \bibnamefont
  {Anderson}},\ }\bibfield  {title} {\bibinfo {title} {{Gauge theory of
  high-temperature superconductors and strongly correlated Fermi systems}},\
  }\href {https://doi.org/10.1103/PhysRevB.37.580} {\bibfield  {journal}
  {\bibinfo  {journal} {Phys. Rev. B}\ }\textbf {\bibinfo {volume} {37}},\
  \bibinfo {pages} {580} (\bibinfo {year} {1988})}\BibitemShut {NoStop}%
\bibitem [{\citenamefont {Baskaran}\ \emph {et~al.}(1987)\citenamefont
  {Baskaran}, \citenamefont {Zou},\ and\ \citenamefont
  {Anderson}}]{Baskaran_1989}%
  \BibitemOpen
  \bibfield  {author} {\bibinfo {author} {\bibfnamefont {G.}~\bibnamefont
  {Baskaran}}, \bibinfo {author} {\bibfnamefont {Z.}~\bibnamefont {Zou}},\ and\
  \bibinfo {author} {\bibfnamefont {P.}~\bibnamefont {Anderson}},\ }\bibfield
  {title} {\bibinfo {title} {{The resonating valence bond state and high-Tc
  superconductivity {\textemdash} A mean field theory}},\ }\href
  {https://doi.org/10.1016/0038-1098(87)90642-9} {\bibfield  {journal}
  {\bibinfo  {journal} {Solid State Communications}\ }\textbf {\bibinfo
  {volume} {63}},\ \bibinfo {pages} {973} (\bibinfo {year} {1987})}\BibitemShut
  {NoStop}%
\bibitem [{footnote1()}]{footnote1}%
  \BibitemOpen
  \bibinfo {note} {In a different context, another example of such occurrence
  on a Kagome lattice is reported in Ref.~\cite{Zhu_2016}.}\BibitemShut {Stop}%
\bibitem [{\citenamefont {Kitaev}\ and\ \citenamefont
  {Preskill}(2006)}]{kitaev2006}%
  \BibitemOpen
  \bibfield  {author} {\bibinfo {author} {\bibfnamefont {A.}~\bibnamefont
  {Kitaev}}\ and\ \bibinfo {author} {\bibfnamefont {J.}~\bibnamefont
  {Preskill}},\ }\bibfield  {title} {\bibinfo {title} {{Topological
  Entanglement Entropy}},\ }\href
  {https://doi.org/10.1103/PhysRevLett.96.110404} {\bibfield  {journal}
  {\bibinfo  {journal} {Phys. Rev. Lett.}\ }\textbf {\bibinfo {volume} {96}},\
  \bibinfo {pages} {110404} (\bibinfo {year} {2006})}\BibitemShut {NoStop}%
\bibitem [{\citenamefont {Levin}\ and\ \citenamefont {Wen}(2006)}]{levin2006}%
  \BibitemOpen
  \bibfield  {author} {\bibinfo {author} {\bibfnamefont {M.}~\bibnamefont
  {Levin}}\ and\ \bibinfo {author} {\bibfnamefont {X.-G.}\ \bibnamefont
  {Wen}},\ }\bibfield  {title} {\bibinfo {title} {{Detecting Topological Order
  in a Ground State Wave Function}},\ }\href
  {https://doi.org/10.1103/PhysRevLett.96.110405} {\bibfield  {journal}
  {\bibinfo  {journal} {Phys. Rev. Lett.}\ }\textbf {\bibinfo {volume} {96}},\
  \bibinfo {pages} {110405} (\bibinfo {year} {2006})}\BibitemShut {NoStop}%
\bibitem [{\citenamefont {White}(1992)}]{white_prl_1992}%
  \BibitemOpen
  \bibfield  {author} {\bibinfo {author} {\bibfnamefont {S.~R.}\ \bibnamefont
  {White}},\ }\bibfield  {title} {\bibinfo {title} {{Density matrix formulation
  for quantum renormalization groups}},\ }\href
  {https://doi.org/10.1103/PhysRevLett.69.2863} {\bibfield  {journal} {\bibinfo
   {journal} {Phys. Rev. Lett.}\ }\textbf {\bibinfo {volume} {69}},\ \bibinfo
  {pages} {2863} (\bibinfo {year} {1992})}\BibitemShut {NoStop}%
\bibitem [{\citenamefont {White}(1993)}]{white_prb_1993}%
  \BibitemOpen
  \bibfield  {author} {\bibinfo {author} {\bibfnamefont {S.~R.}\ \bibnamefont
  {White}},\ }\bibfield  {title} {\bibinfo {title} {{Density-matrix algorithms
  for quantum renormalization groups}},\ }\href
  {https://doi.org/10.1103/PhysRevB.48.10345} {\bibfield  {journal} {\bibinfo
  {journal} {Phys. Rev. B}\ }\textbf {\bibinfo {volume} {48}},\ \bibinfo
  {pages} {10345} (\bibinfo {year} {1993})}\BibitemShut {NoStop}%
\bibitem [{\citenamefont {Schollw\"{o}ck}(2011)}]{schollwock_aop_2011}%
  \BibitemOpen
  \bibfield  {author} {\bibinfo {author} {\bibfnamefont {U.}~\bibnamefont
  {Schollw\"{o}ck}},\ }\bibfield  {title} {\bibinfo {title} {{The
  density-matrix renormalization group in the age of matrix product states}},\
  }\href {https://doi.org/10.1016/j.aop.2010.09.012} {\bibfield  {journal}
  {\bibinfo  {journal} {Annals of Physics}\ }\textbf {\bibinfo {volume}
  {326}},\ \bibinfo {pages} {96} (\bibinfo {year} {2011})}\BibitemShut
  {NoStop}%
\bibitem [{\citenamefont {Or{\'{u}}s}(2014)}]{Orus_aop_2014}%
  \BibitemOpen
  \bibfield  {author} {\bibinfo {author} {\bibfnamefont {R.}~\bibnamefont
  {Or{\'{u}}s}},\ }\bibfield  {title} {\bibinfo {title} {{A practical
  introduction to tensor networks: Matrix product states and projected
  entangled pair states}},\ }\href {https://doi.org/10.1016/j.aop.2014.06.013}
  {\bibfield  {journal} {\bibinfo  {journal} {Annals of Physics}\ }\textbf
  {\bibinfo {volume} {349}},\ \bibinfo {pages} {117} (\bibinfo {year}
  {2014})}\BibitemShut {NoStop}%
\bibitem [{footnote5()}]{footnote5}%
  \BibitemOpen
  \bibinfo {note} {As we report in the supplemental material \cite{supmat}, an
  extra intermediate phase occurs for $0.25 \lesssim g \lesssim
  0.4$.}\BibitemShut {Stop}%
\bibitem [{\citenamefont {Dodds}\ \emph {et~al.}(2013)\citenamefont {Dodds},
  \citenamefont {Bhattacharjee},\ and\ \citenamefont {Kim}}]{Dodds_2013}%
  \BibitemOpen
  \bibfield  {author} {\bibinfo {author} {\bibfnamefont {T.}~\bibnamefont
  {Dodds}}, \bibinfo {author} {\bibfnamefont {S.}~\bibnamefont
  {Bhattacharjee}},\ and\ \bibinfo {author} {\bibfnamefont {Y.~B.}\
  \bibnamefont {Kim}},\ }\bibfield  {title} {\bibinfo {title} {{Quantum spin
  liquids in the absence of spin-rotation symmetry: Application to
  herbertsmithite}},\ }\href {https://doi.org/10.1103/PhysRevB.88.224413}
  {\bibfield  {journal} {\bibinfo  {journal} {Phys. Rev. B}\ }\textbf {\bibinfo
  {volume} {88}},\ \bibinfo {pages} {224413} (\bibinfo {year}
  {2013})}\BibitemShut {NoStop}%
\bibitem [{\citenamefont {Reuther}\ \emph {et~al.}(2014)\citenamefont
  {Reuther}, \citenamefont {Lee},\ and\ \citenamefont {Alicea}}]{Reuther_2014}%
  \BibitemOpen
  \bibfield  {author} {\bibinfo {author} {\bibfnamefont {J.}~\bibnamefont
  {Reuther}}, \bibinfo {author} {\bibfnamefont {S.-P.}\ \bibnamefont {Lee}},\
  and\ \bibinfo {author} {\bibfnamefont {J.}~\bibnamefont {Alicea}},\
  }\bibfield  {title} {\bibinfo {title} {{Classification of spin liquids on the
  square lattice with strong spin-orbit coupling}},\ }\href
  {https://doi.org/10.1103/PhysRevB.90.174417} {\bibfield  {journal} {\bibinfo
  {journal} {Phys. Rev. B}\ }\textbf {\bibinfo {volume} {90}},\ \bibinfo
  {pages} {174417} (\bibinfo {year} {2014})}\BibitemShut {NoStop}%
\bibitem [{footnote2()}]{footnote2}%
  \BibitemOpen
  \bibinfo {note} {The ``singlet'' and ``triplet'' terminology is derived from
  the discussion of SU(2) spin-rotation symmetric Hamiltonians, which have been
  extensively studied in the literature. If the state does not spontaneously
  break the spin-rotation symmetry, only singlet terms are present. However, in
  the presence of spin-rotation symmetry-breaking perturbations, both terms may
  be present, and more generally the mean-field Hamiltonian has to be written
  in terms of a four-component spinor $\Psi=(f_{\uparrow} \ \
  f_{\downarrow}^\dagger \ \ f_{\downarrow} \ \ -f_{\uparrow}^\dagger)^T$. If
  the spin-rotation is only broken down to U(1), it is still possible to use a
  two-component spinor representation as in Eq. \eqref{eq:mf_Ham}, where the
  singlet and triplet terms can be present without mixing.}\BibitemShut {Stop}%
\bibitem [{\citenamefont {Wen}(2007)}]{wen2004}%
  \BibitemOpen
  \bibfield  {author} {\bibinfo {author} {\bibfnamefont {X.-G.}\ \bibnamefont
  {Wen}},\ }\href {https://doi.org/10.1093/acprof:oso/9780199227259.001.0001}
  {\emph {\bibinfo {title} {{Quantum Field Theory of Many-Body Systems}}}}\
  (\bibinfo  {publisher} {Oxford University Press},\ \bibinfo {year}
  {2007})\BibitemShut {NoStop}%
\bibitem [{\citenamefont {Wen}\ \emph {et~al.}(1989)\citenamefont {Wen},
  \citenamefont {Wilczek},\ and\ \citenamefont {Zee}}]{Wen_1989}%
  \BibitemOpen
  \bibfield  {author} {\bibinfo {author} {\bibfnamefont {X.~G.}\ \bibnamefont
  {Wen}}, \bibinfo {author} {\bibfnamefont {F.}~\bibnamefont {Wilczek}},\ and\
  \bibinfo {author} {\bibfnamefont {A.}~\bibnamefont {Zee}},\ }\bibfield
  {title} {\bibinfo {title} {{Chiral spin states and superconductivity}},\
  }\href {https://doi.org/10.1103/PhysRevB.39.11413} {\bibfield  {journal}
  {\bibinfo  {journal} {Phys. Rev. B}\ }\textbf {\bibinfo {volume} {39}},\
  \bibinfo {pages} {11413} (\bibinfo {year} {1989})}\BibitemShut {NoStop}%
\bibitem [{\citenamefont {Lu}\ and\ \citenamefont {Ran}(2011)}]{Lu_2011}%
  \BibitemOpen
  \bibfield  {author} {\bibinfo {author} {\bibfnamefont {Y.-M.}\ \bibnamefont
  {Lu}}\ and\ \bibinfo {author} {\bibfnamefont {Y.}~\bibnamefont {Ran}},\
  }\bibfield  {title} {\bibinfo {title} {{${\mathbb{Z}}_{2}$ spin liquid and
  chiral antiferromagnetic phase in the Hubbard model on a honeycomb
  lattice}},\ }\href {https://doi.org/10.1103/PhysRevB.84.024420} {\bibfield
  {journal} {\bibinfo  {journal} {Phys. Rev. B}\ }\textbf {\bibinfo {volume}
  {84}},\ \bibinfo {pages} {024420} (\bibinfo {year} {2011})}\BibitemShut
  {NoStop}%
\bibitem [{footnote3()}]{footnote3}%
  \BibitemOpen
  \bibinfo {note} {We include all terms that are allowed by symmetry for each
  link, which in principle can be present in a mean-field state. However, if
  the mean-field Hamiltonian is restricted in the range of interactions (such
  as up to NNN interactions in our case), some of the terms can be removed by a
  gauge transformation.}\BibitemShut {Stop}%
\bibitem [{\citenamefont {Hermele}(2007)}]{Hermele_2007}%
  \BibitemOpen
  \bibfield  {author} {\bibinfo {author} {\bibfnamefont {M.}~\bibnamefont
  {Hermele}},\ }\bibfield  {title} {\bibinfo {title} {{SU(2) gauge theory of
  the Hubbard model and application to the honeycomb lattice}},\ }\href
  {https://doi.org/10.1103/PhysRevB.76.035125} {\bibfield  {journal} {\bibinfo
  {journal} {Phys. Rev. B}\ }\textbf {\bibinfo {volume} {76}},\ \bibinfo
  {pages} {035125} (\bibinfo {year} {2007})}\BibitemShut {NoStop}%
\bibitem [{\citenamefont {Zhang}\ \emph {et~al.}(2011)\citenamefont {Zhang},
  \citenamefont {Grover},\ and\ \citenamefont {Vishwanath}}]{Zhang_2011}%
  \BibitemOpen
  \bibfield  {author} {\bibinfo {author} {\bibfnamefont {Y.}~\bibnamefont
  {Zhang}}, \bibinfo {author} {\bibfnamefont {T.}~\bibnamefont {Grover}},\ and\
  \bibinfo {author} {\bibfnamefont {A.}~\bibnamefont {Vishwanath}},\ }\bibfield
   {title} {\bibinfo {title} {{Topological entanglement entropy of
  ${\mathbb{Z}}_{2}$ spin liquids and lattice Laughlin states}},\ }\href
  {https://doi.org/10.1103/PhysRevB.84.075128} {\bibfield  {journal} {\bibinfo
  {journal} {Phys. Rev. B}\ }\textbf {\bibinfo {volume} {84}},\ \bibinfo
  {pages} {075128} (\bibinfo {year} {2011})}\BibitemShut {NoStop}%
\bibitem [{\citenamefont {Zhang}\ and\ \citenamefont
  {Vishwanath}(2013)}]{Zhang_2013}%
  \BibitemOpen
  \bibfield  {author} {\bibinfo {author} {\bibfnamefont {Y.}~\bibnamefont
  {Zhang}}\ and\ \bibinfo {author} {\bibfnamefont {A.}~\bibnamefont
  {Vishwanath}},\ }\bibfield  {title} {\bibinfo {title} {{Establishing
  non-Abelian topological order in Gutzwiller-projected Chern insulators via
  entanglement entropy and modular $\mathcal{S}$-matrix}},\ }\href
  {https://doi.org/10.1103/PhysRevB.87.161113} {\bibfield  {journal} {\bibinfo
  {journal} {Phys. Rev. B}\ }\textbf {\bibinfo {volume} {87}},\ \bibinfo
  {pages} {161113} (\bibinfo {year} {2013})}\BibitemShut {NoStop}%
\bibitem [{\citenamefont {Laughlin}(1983)}]{laughlin1983}%
  \BibitemOpen
  \bibfield  {author} {\bibinfo {author} {\bibfnamefont {R.~B.}\ \bibnamefont
  {Laughlin}},\ }\bibfield  {title} {\bibinfo {title} {{Anomalous Quantum Hall
  Effect: An Incompressible Quantum Fluid with Fractionally Charged
  Excitations}},\ }\href {https://doi.org/10.1103/PhysRevLett.50.1395}
  {\bibfield  {journal} {\bibinfo  {journal} {Phys. Rev. Lett.}\ }\textbf
  {\bibinfo {volume} {50}},\ \bibinfo {pages} {1395} (\bibinfo {year}
  {1983})}\BibitemShut {NoStop}%
\bibitem [{\citenamefont {Mei}\ and\ \citenamefont {Wen}(2015)}]{Wei_2015}%
  \BibitemOpen
  \bibfield  {author} {\bibinfo {author} {\bibfnamefont {J.-W.}\ \bibnamefont
  {Mei}}\ and\ \bibinfo {author} {\bibfnamefont {X.-G.}\ \bibnamefont {Wen}},\
  }\bibfield  {title} {\bibinfo {title} {{Modular matrices from universal
  wave-function overlaps in Gutzwiller-projected parton wave functions}},\
  }\href {https://doi.org/10.1103/PhysRevB.91.125123} {\bibfield  {journal}
  {\bibinfo  {journal} {Phys. Rev. B}\ }\textbf {\bibinfo {volume} {91}},\
  \bibinfo {pages} {125123} (\bibinfo {year} {2015})}\BibitemShut {NoStop}%
\bibitem [{\citenamefont {Thouless}(1998)}]{thouless1998topological}%
  \BibitemOpen
  \bibfield  {author} {\bibinfo {author} {\bibfnamefont {D.}~\bibnamefont
  {Thouless}},\ }\href@noop {} {\emph {\bibinfo {title} {Topological quantum
  numbers in nonrelativistic physics}}}\ (\bibinfo  {publisher} {World
  Scientific},\ \bibinfo {year} {1998})\BibitemShut {NoStop}%
\bibitem [{\citenamefont {Wietek}\ \emph {et~al.}(2015)\citenamefont {Wietek},
  \citenamefont {Sterdyniak},\ and\ \citenamefont {L\"auchli}}]{Wietek_2015}%
  \BibitemOpen
  \bibfield  {author} {\bibinfo {author} {\bibfnamefont {A.}~\bibnamefont
  {Wietek}}, \bibinfo {author} {\bibfnamefont {A.}~\bibnamefont {Sterdyniak}},\
  and\ \bibinfo {author} {\bibfnamefont {A.~M.}\ \bibnamefont {L\"auchli}},\
  }\bibfield  {title} {\bibinfo {title} {{Nature of chiral spin liquids on the
  kagome lattice}},\ }\href {https://doi.org/10.1103/PhysRevB.92.125122}
  {\bibfield  {journal} {\bibinfo  {journal} {Phys. Rev. B}\ }\textbf {\bibinfo
  {volume} {92}},\ \bibinfo {pages} {125122} (\bibinfo {year}
  {2015})}\BibitemShut {NoStop}%
\bibitem [{\citenamefont {Wietek}\ and\ \citenamefont
  {L\"auchli}(2017)}]{Wietek_2017}%
  \BibitemOpen
  \bibfield  {author} {\bibinfo {author} {\bibfnamefont {A.}~\bibnamefont
  {Wietek}}\ and\ \bibinfo {author} {\bibfnamefont {A.~M.}\ \bibnamefont
  {L\"auchli}},\ }\bibfield  {title} {\bibinfo {title} {{Chiral spin liquid and
  quantum criticality in extended $S=\frac{1}{2}$ Heisenberg models on the
  triangular lattice}},\ }\href {https://doi.org/10.1103/PhysRevB.95.035141}
  {\bibfield  {journal} {\bibinfo  {journal} {Phys. Rev. B}\ }\textbf {\bibinfo
  {volume} {95}},\ \bibinfo {pages} {035141} (\bibinfo {year}
  {2017})}\BibitemShut {NoStop}%
\bibitem [{\citenamefont {Wang}\ \emph {et~al.}(2011)\citenamefont {Wang},
  \citenamefont {Gu}, \citenamefont {Gong},\ and\ \citenamefont
  {Sheng}}]{wang2011}%
  \BibitemOpen
  \bibfield  {author} {\bibinfo {author} {\bibfnamefont {Y.-F.}\ \bibnamefont
  {Wang}}, \bibinfo {author} {\bibfnamefont {Z.-C.}\ \bibnamefont {Gu}},
  \bibinfo {author} {\bibfnamefont {C.-D.}\ \bibnamefont {Gong}},\ and\
  \bibinfo {author} {\bibfnamefont {D.~N.}\ \bibnamefont {Sheng}},\ }\bibfield
  {title} {\bibinfo {title} {{Fractional Quantum Hall Effect of Hard-Core
  Bosons in Topological Flat Bands}},\ }\href
  {https://doi.org/10.1103/PhysRevLett.107.146803} {\bibfield  {journal}
  {\bibinfo  {journal} {Phys. Rev. Lett.}\ }\textbf {\bibinfo {volume} {107}},\
  \bibinfo {pages} {146803} (\bibinfo {year} {2011})}\BibitemShut {NoStop}%
\bibitem [{\citenamefont {He}\ \emph {et~al.}(2017)\citenamefont {He},
  \citenamefont {Zaletel}, \citenamefont {Oshikawa},\ and\ \citenamefont
  {Pollmann}}]{he2017}%
  \BibitemOpen
  \bibfield  {author} {\bibinfo {author} {\bibfnamefont {Y.-C.}\ \bibnamefont
  {He}}, \bibinfo {author} {\bibfnamefont {M.~P.}\ \bibnamefont {Zaletel}},
  \bibinfo {author} {\bibfnamefont {M.}~\bibnamefont {Oshikawa}},\ and\
  \bibinfo {author} {\bibfnamefont {F.}~\bibnamefont {Pollmann}},\ }\bibfield
  {title} {\bibinfo {title} {{Signatures of Dirac Cones in a DMRG Study of the
  Kagome Heisenberg Model}},\ }\href
  {https://doi.org/10.1103/PhysRevX.7.031020} {\bibfield  {journal} {\bibinfo
  {journal} {Phys. Rev. X}\ }\textbf {\bibinfo {volume} {7}},\ \bibinfo {pages}
  {031020} (\bibinfo {year} {2017})}\BibitemShut {NoStop}%
\bibitem [{\citenamefont {Hu}\ \emph {et~al.}(2019)\citenamefont {Hu},
  \citenamefont {Zhu}, \citenamefont {Eggert},\ and\ \citenamefont
  {He}}]{hu2019}%
  \BibitemOpen
  \bibfield  {author} {\bibinfo {author} {\bibfnamefont {S.}~\bibnamefont
  {Hu}}, \bibinfo {author} {\bibfnamefont {W.}~\bibnamefont {Zhu}}, \bibinfo
  {author} {\bibfnamefont {S.}~\bibnamefont {Eggert}},\ and\ \bibinfo {author}
  {\bibfnamefont {Y.-C.}\ \bibnamefont {He}},\ }\bibfield  {title} {\bibinfo
  {title} {{Dirac Spin Liquid on the Spin-$1/2$ Triangular Heisenberg
  Antiferromagnet}},\ }\href {https://doi.org/10.1103/PhysRevLett.123.207203}
  {\bibfield  {journal} {\bibinfo  {journal} {Phys. Rev. Lett.}\ }\textbf
  {\bibinfo {volume} {123}},\ \bibinfo {pages} {207203} (\bibinfo {year}
  {2019})}\BibitemShut {NoStop}%
\bibitem [{\citenamefont {Ferrari}\ \emph {et~al.}(2021)\citenamefont
  {Ferrari}, \citenamefont {Parola},\ and\ \citenamefont
  {Becca}}]{ferrari2021}%
  \BibitemOpen
  \bibfield  {author} {\bibinfo {author} {\bibfnamefont {F.}~\bibnamefont
  {Ferrari}}, \bibinfo {author} {\bibfnamefont {A.}~\bibnamefont {Parola}},\
  and\ \bibinfo {author} {\bibfnamefont {F.}~\bibnamefont {Becca}},\ }\bibfield
   {title} {\bibinfo {title} {{Gapless spin liquids in disguise}},\ }\href
  {https://doi.org/10.1103/PhysRevB.103.195140} {\bibfield  {journal} {\bibinfo
   {journal} {Phys. Rev. B}\ }\textbf {\bibinfo {volume} {103}},\ \bibinfo
  {pages} {195140} (\bibinfo {year} {2021})}\BibitemShut {NoStop}%
\bibitem [{footnote6()}]{footnote6}%
  \BibitemOpen
  \bibinfo {note} {We, however, note that the finite volume effects prevent a
  clear interpretation of this data as a striking signature of topological
  order: such effects are expected given the observed strong finite-size
  effects in the numerical simulations.}\BibitemShut {Stop}%
\bibitem [{footnote4()}]{footnote4}%
  \BibitemOpen
  \bibinfo {note} {We consider two different cylinder geometries for our
  calculations, namely the geometries I and II, as demarcated
  in~\cite{pollmann_prl_2015}.}\BibitemShut {Stop}%
\bibitem [{\citenamefont {Gu}\ and\ \citenamefont {Lin}(2009)}]{Gu2009}%
  \BibitemOpen
  \bibfield  {author} {\bibinfo {author} {\bibfnamefont {S.-J.}\ \bibnamefont
  {Gu}}\ and\ \bibinfo {author} {\bibfnamefont {H.-Q.}\ \bibnamefont {Lin}},\
  }\bibfield  {title} {\bibinfo {title} {{Scaling dimension of fidelity
  susceptibility in quantum phase transitions}},\ }\href
  {https://doi.org/10.1209/0295-5075/87/10003} {\bibfield  {journal} {\bibinfo
  {journal} {{EPL} (Europhysics Letters)}\ }\textbf {\bibinfo {volume} {87}},\
  \bibinfo {pages} {10003} (\bibinfo {year} {2009})}\BibitemShut {NoStop}%
\bibitem [{\citenamefont {Wen}(2013)}]{Wen2013}%
  \BibitemOpen
  \bibfield  {author} {\bibinfo {author} {\bibfnamefont {X.-G.}\ \bibnamefont
  {Wen}},\ }\bibfield  {title} {\bibinfo {title} {{Topological Order: From
  Long-Range Entangled Quantum Matter to a Unified Origin of Light and
  Electrons}},\ }\href {https://doi.org/10.1155/2013/198710} {\bibfield
  {journal} {\bibinfo  {journal} {{ISRN} Condensed Matter Physics}\ }\textbf
  {\bibinfo {volume} {2013}},\ \bibinfo {pages} {1} (\bibinfo {year}
  {2013})}\BibitemShut {NoStop}%
\bibitem [{\citenamefont {Barredo}\ \emph {et~al.}(2018)\citenamefont
  {Barredo}, \citenamefont {Lienhard}, \citenamefont {de~L{\'{e}}s{\'{e}}leuc},
  \citenamefont {Lahaye},\ and\ \citenamefont
  {Browaeys}}]{barredo2018synthetic}%
  \BibitemOpen
  \bibfield  {author} {\bibinfo {author} {\bibfnamefont {D.}~\bibnamefont
  {Barredo}}, \bibinfo {author} {\bibfnamefont {V.}~\bibnamefont {Lienhard}},
  \bibinfo {author} {\bibfnamefont {S.}~\bibnamefont
  {de~L{\'{e}}s{\'{e}}leuc}}, \bibinfo {author} {\bibfnamefont
  {T.}~\bibnamefont {Lahaye}},\ and\ \bibinfo {author} {\bibfnamefont
  {A.}~\bibnamefont {Browaeys}},\ }\bibfield  {title} {\bibinfo {title}
  {{Synthetic three-dimensional atomic structures assembled atom by atom}},\
  }\href {https://doi.org/10.1038/s41586-018-0450-2} {\bibfield  {journal}
  {\bibinfo  {journal} {Nature}\ }\textbf {\bibinfo {volume} {561}},\ \bibinfo
  {pages} {79} (\bibinfo {year} {2018})}\BibitemShut {NoStop}%
\bibitem [{\citenamefont {Fishman}\ \emph {et~al.}(2022)\citenamefont
  {Fishman}, \citenamefont {White},\ and\ \citenamefont
  {Stoudenmire}}]{itensor}%
  \BibitemOpen
  \bibfield  {author} {\bibinfo {author} {\bibfnamefont {M.}~\bibnamefont
  {Fishman}}, \bibinfo {author} {\bibfnamefont {S.~R.}\ \bibnamefont {White}},\
  and\ \bibinfo {author} {\bibfnamefont {E.~M.}\ \bibnamefont {Stoudenmire}},\
  }\bibfield  {title} {\bibinfo {title} {{The ITensor Software Library for
  Tensor Network Calculations}},\ }\href
  {https://doi.org/10.21468/SciPostPhysCodeb.4} {\bibfield  {journal} {\bibinfo
   {journal} {SciPost Phys. Codebases}\ ,\ \bibinfo {pages} {4}} (\bibinfo
  {year} {2022})}\BibitemShut {NoStop}%
\bibitem [{\citenamefont {Hauschild}\ and\ \citenamefont
  {Pollmann}(2018)}]{tenpy}%
  \BibitemOpen
  \bibfield  {author} {\bibinfo {author} {\bibfnamefont {J.}~\bibnamefont
  {Hauschild}}\ and\ \bibinfo {author} {\bibfnamefont {F.}~\bibnamefont
  {Pollmann}},\ }\bibfield  {title} {\bibinfo {title} {{Efficient numerical
  simulations with Tensor Networks: Tensor Network Python (TeNPy)}},\ }\href
  {https://doi.org/10.21468/SciPostPhysLectNotes.5} {\bibfield  {journal}
  {\bibinfo  {journal} {SciPost Phys. Lect. Notes}\ ,\ \bibinfo {pages} {5}}
  (\bibinfo {year} {2018})},\ \bibinfo {note} {code available from
  \url{https://github.com/tenpy/tenpy}},\ \Eprint
  {https://arxiv.org/abs/1805.00055} {arXiv:1805.00055} \BibitemShut {NoStop}%
\bibitem [{\citenamefont {Vidal}(2007)}]{Vidal_2007}%
  \BibitemOpen
  \bibfield  {author} {\bibinfo {author} {\bibfnamefont {G.}~\bibnamefont
  {Vidal}},\ }\bibfield  {title} {\bibinfo {title} {{Classical Simulation of
  Infinite-Size Quantum Lattice Systems in One Spatial Dimension}},\ }\href
  {https://doi.org/10.1103/PhysRevLett.98.070201} {\bibfield  {journal}
  {\bibinfo  {journal} {Phys. Rev. Lett.}\ }\textbf {\bibinfo {volume} {98}},\
  \bibinfo {pages} {070201} (\bibinfo {year} {2007})}\BibitemShut {NoStop}%
\bibitem [{\citenamefont {Kj\"all}\ \emph {et~al.}(2013)\citenamefont
  {Kj\"all}, \citenamefont {Zaletel}, \citenamefont {Mong}, \citenamefont
  {Bardarson},\ and\ \citenamefont {Pollmann}}]{Kjall_PRB_2013}%
  \BibitemOpen
  \bibfield  {author} {\bibinfo {author} {\bibfnamefont {J.~A.}\ \bibnamefont
  {Kj\"all}}, \bibinfo {author} {\bibfnamefont {M.~P.}\ \bibnamefont
  {Zaletel}}, \bibinfo {author} {\bibfnamefont {R.~S.~K.}\ \bibnamefont
  {Mong}}, \bibinfo {author} {\bibfnamefont {J.~H.}\ \bibnamefont
  {Bardarson}},\ and\ \bibinfo {author} {\bibfnamefont {F.}~\bibnamefont
  {Pollmann}},\ }\bibfield  {title} {\bibinfo {title} {{Phase diagram of the
  anisotropic spin-2 XXZ model: Infinite-system density matrix renormalization
  group study}},\ }\href {https://doi.org/10.1103/PhysRevB.87.235106}
  {\bibfield  {journal} {\bibinfo  {journal} {Phys. Rev. B}\ }\textbf {\bibinfo
  {volume} {87}},\ \bibinfo {pages} {235106} (\bibinfo {year}
  {2013})}\BibitemShut {NoStop}%
\bibitem [{\citenamefont {McCulloch}(2008)}]{McCulloch_2008}%
  \BibitemOpen
  \bibfield  {author} {\bibinfo {author} {\bibfnamefont {I.~P.}\ \bibnamefont
  {McCulloch}},\ }\href@noop {} {\bibinfo {title} {{Infinite size density
  matrix renormalization group, revisited}}} (\bibinfo {year} {2008}),\ \Eprint
  {https://arxiv.org/abs/arXiv:0804.2509} {arXiv:0804.2509} \BibitemShut
  {NoStop}%
\bibitem [{\citenamefont {Crosswhite}\ \emph {et~al.}(2008)\citenamefont
  {Crosswhite}, \citenamefont {Doherty},\ and\ \citenamefont
  {Vidal}}]{Crosswhite_2008}%
  \BibitemOpen
  \bibfield  {author} {\bibinfo {author} {\bibfnamefont {G.~M.}\ \bibnamefont
  {Crosswhite}}, \bibinfo {author} {\bibfnamefont {A.~C.}\ \bibnamefont
  {Doherty}},\ and\ \bibinfo {author} {\bibfnamefont {G.}~\bibnamefont
  {Vidal}},\ }\bibfield  {title} {\bibinfo {title} {{Applying matrix product
  operators to model systems with long-range interactions}},\ }\href
  {https://doi.org/10.1103/PhysRevB.78.035116} {\bibfield  {journal} {\bibinfo
  {journal} {Phys. Rev. B}\ }\textbf {\bibinfo {volume} {78}},\ \bibinfo
  {pages} {035116} (\bibinfo {year} {2008})}\BibitemShut {NoStop}%
\bibitem [{\citenamefont {He}\ \emph {et~al.}(2015)\citenamefont {He},
  \citenamefont {Bhattacharjee}, \citenamefont {Moessner},\ and\ \citenamefont
  {Pollmann}}]{pollmann_prl_2015}%
  \BibitemOpen
  \bibfield  {author} {\bibinfo {author} {\bibfnamefont {Y.-C.}\ \bibnamefont
  {He}}, \bibinfo {author} {\bibfnamefont {S.}~\bibnamefont {Bhattacharjee}},
  \bibinfo {author} {\bibfnamefont {R.}~\bibnamefont {Moessner}},\ and\
  \bibinfo {author} {\bibfnamefont {F.}~\bibnamefont {Pollmann}},\ }\bibfield
  {title} {\bibinfo {title} {{Bosonic Integer Quantum Hall Effect in an
  Interacting Lattice Model}},\ }\href
  {https://doi.org/10.1103/PhysRevLett.115.116803} {\bibfield  {journal}
  {\bibinfo  {journal} {Phys. Rev. Lett.}\ }\textbf {\bibinfo {volume} {115}},\
  \bibinfo {pages} {116803} (\bibinfo {year} {2015})}\BibitemShut {NoStop}%
\bibitem [{\citenamefont {Zhu}\ \emph {et~al.}(2016)\citenamefont {Zhu},
  \citenamefont {Gong},\ and\ \citenamefont {Sheng}}]{Zhu_2016}%
  \BibitemOpen
  \bibfield  {author} {\bibinfo {author} {\bibfnamefont {W.}~\bibnamefont
  {Zhu}}, \bibinfo {author} {\bibfnamefont {S.~S.}\ \bibnamefont {Gong}},\ and\
  \bibinfo {author} {\bibfnamefont {D.~N.}\ \bibnamefont {Sheng}},\ }\bibfield
  {title} {\bibinfo {title} {{Interaction-driven fractional quantum Hall state
  of hard-core bosons on kagome lattice at one-third filling}},\ }\href
  {https://doi.org/10.1103/PhysRevB.94.035129} {\bibfield  {journal} {\bibinfo
  {journal} {Phys. Rev. B}\ }\textbf {\bibinfo {volume} {94}},\ \bibinfo
  {pages} {035129} (\bibinfo {year} {2016})}\BibitemShut {NoStop}%
\end{thebibliography}%

\end{document}